\documentclass[UTF8,prc,twocolumn,showpacs,preprintnumbers,amsmath,amssymb]{revtex4-1}
\usepackage[T1]{fontenc}
\usepackage{CJK}
\usepackage{tipa}
\usepackage{graphicx}
\usepackage{dcolumn}
\usepackage{bm}
\usepackage{amssymb}
\usepackage{multirow}
\usepackage{booktabs}
\usepackage[section]{placeins}
\usepackage{textcomp}
\usepackage{color,xcolor}
\usepackage{lineno}
\usepackage[driverfallback=dvipdfm,pdfstartview=FitH,CJKbookmarks=true,bookmarksnumbered=true,bookmarksopen=true,colorlinks=true,pdfborder=001,linkcolor=blue,anchorcolor=green,urlcolor=blue,citecolor=blue]{hyperref}

\begin{document}
\begin{CJK*}{UTF8}{song}

\title{Simultaneous measurement of beta-delayed proton and gamma emission of $^{26}$P for $^{25}$Al($p,\gamma$)$^{26}$Si reaction rate}

\author{P. F. Liang$^1$}
\altaffiliation{These authors contributed equally to this work and should be considered as co-first authors.\label{Co-authors}}
\author{L. J. Sun$^{2,3,}$\textsuperscript{\ref {Co-authors}}}
\author{J. Lee$^1$ }
\email{jleehc@hku.hk}
\author{S. Q. Hou$^{4}$}
\author{X. X. Xu$^{1,2,4}$}
\email{xinxing@impcas.ac.cn}
\author{C. J. Lin$^{2,5}$}
\email{cjlin@ciae.ac.cn}
\author{C. X. Yuan$^{6}$}
\author{J. J. He$^{7,8}$}
\author{\\Z. H. Li$^{9}$}
\author{J. S. Wang$^{10,4,8}$}
\author{D. X. Wang$^{2}$}
\author{H. Y. Wu$^{9}$}
\author{Y. Y. Yang$^{4}$}
\author{Y. H. Lam$^{4}$}
\author{P. Ma$^{4}$}
\author{F. F. Duan$^{11,4}$}
\author{\\Z. H. Gao$^{4,11}$}
\author{Q. Hu$^{4}$}
\author{Z. Bai$^{4}$}
\author{J. B. Ma$^{4}$}
\author{J. G. Wang$^{4}$}
\author{F. P. Zhong$^{5,2}$}
\author{C. G. Wu$^{9}$}
\author{D. W. Luo$^{9}$}
\author{Y. Jiang$^{9}$}
\author{\\Y. Liu$^{9}$}
\author{D. S. Hou$^{4,8}$}
\author{R. Li$^{4,8}$}
\author{N. R. Ma$^{2}$}
\author{W. H. Ma$^{4,12}$}
\author{G. Z. Shi$^{4}$}
\author{G. M. Yu$^{4}$}
\author{D. Patel$^{4}$}
\author{S. Y. Jin$^{4,8}$}
\author{\\Y. F. Wang$^{13,4}$}
\author{Y. C. Yu$^{13,4}$}
\author{Q. W. Zhou$^{14,4}$}
\author{P. Wang$^{14,4}$}
\author{L. Y. Hu$^{15}$}
\author{X. Wang$^{9}$}
\author{H. L. Zang$^{9}$}
\author{P. J. Li$^{1}$}
\author{\\Q. Q. Zhao$^{1}$}
\author{L. Yang$^{2}$}
\author{P. W. Wen$^{2}$}
\author{F. Yang$^{2}$}
\author{H. M. Jia$^{2}$}
\author{G. L. Zhang$^{16}$}
\author{M. Pan$^{16,2}$}
\author{X. Y. Wang$^{16}$}
\author{\\H. H. Sun$^{2}$}
\author{Z. G. Hu$^{4}$}
\author{R. F. Chen$^{4}$}
\author{M. L. Liu$^{4}$}
\author{W. Q. Yang$^{4}$}
\author{Y. M. Zhao$^{3}$}
\affiliation{\text{$^1$Department of Physics, The University of Hong Kong, Hong Kong, China}\\
	\text{$^2$Department of Nuclear Physics, China Institute of Atomic Energy, Beijing 102413, China}\\
	\text{$^3$School of Physics and Astronomy, Shanghai Jiao Tong University, Shanghai 200240, China}\\
	\text{$^4$Institute of Modern Physics, Chinese Academy of Sciences, Lanzhou 730000, China}\\
	\text{$^5$College of Physics and Technology, Guangxi Normal University, Guilin 541004, China}\\
	\text{$^6$Sino-French Institute of Nuclear Engineering and Technology, Sun Yat-Sen University, Zhuhai 519082, China}\\
	\text{$^7$College of Nuclear Science and Technology, Beijing Normal University, Beijing 100875, China}\\
	\text{$^8$University of Chinese Academy of Sciences, Beijing 100049, China}\\
	\text{$^9$State Key Laboratory of Nuclear Physics and Technology, School of Physics, Peking University, Beijing 100871, China}\\
	\text{$^{10}$School of Science, Huzhou University, Huzhou 313000, China}
	\text{$^{11}$School of Nuclear Science and Technology, Lanzhou University, Lanzhou 730000, China}\\
	\text{$^{12}$Institute of Modern Physics, Fudan University, Shanghai 200433, China}\\
	\text{$^{13}$School of Physics and Astronomy, Yunnan University, Kunming 650091, China}\\
	\text{$^{14}$School of Physical Science and Technology, Southwest University, Chongqing 400044, China}\\
	$^{15}$Fundamental Science on Nuclear Safety and Simulation Technology Laboratory,
		\\Harbin Engineering University, Harbin 150001, China\\
	\text{$^{16}$School of Physics and Nuclear Energy Engineering, Beihang University, Beijing 100191, China}}%
\date{\today}

\begin{abstract} 
	$\beta$ decay of $^{26}$P was used to populate the astrophysically important $E_x=$5929.4(8) keV $J^{\pi}=3{^+}$ state of $^{26}$Si. Both $\beta$-delayed proton at 418(8) keV and gamma ray at 1742(2) keV emitted from this state were measured simultaneously for the first time with corresponding absolute intensities of 11.1(12)\% and 0.59(44)\%, respectively. Besides, shell model calculations with weakly bound effects were performed to investigate the decay properties of other resonant states and a spin-parity of $4^+$ rather than $0^+$ was favored for the $E_x=$5945.9(40) keV state. Combining the experimental results and theoretical calculations, $^{25}$Al($p,\gamma$)$^{26}$Si reaction rate in explosive hydrogen burning environments was calculated and compared with previous studies.  

\end{abstract}

\pacs{Valid PACS appear here}

\maketitle

\end{CJK*}

\section{\label{sec:level1} introduction }

As the half-life of $^{26}$Al ($T_{1/2}=7.17\times10^{5}$ y) is much less than the age of the Galaxy ($\approx10^{10}$ y), the observation of 1809 keV $\gamma$ ray from $\beta$ decay of $^{26}$Al could directly prove that the stellar nucleosynthesis processes are currently active in our Galaxy. In previous satellite-based astronomical observations, the mass of Galactic $^{26}$Al was estimated to be 2.7$ \pm $0.7 solar masses ($M_{\odot} $)\cite{Wang2009, Diehl2006}. The primary sites of Galactic $^{26}$Al was suggested to be massive stars and core-collapse supernovae concluded from the spatial distribution of $^{26}$Al\cite{Knodlseder1999,Pluschke2001}. However, the observed $^{60}$Fe/$^{26}$Al $\gamma$-ray flux ratio\cite{Smith2004,Harris2005} was smaller than theoretical predictions which indicated that there should be other important sources for Galactic $^{26}$Al\cite{Rauscher2002,Woosley2007}. As one of the most frequent types of thermonuclear stellar explosions in the Galaxy, classical novae were expected to contribute 0.1-0.4 $M_{\odot}$ of Galactic $^{26}$Al\cite{Parikh2014,Jose1997}, or even up to 0.6 $M_{\odot}$\cite{Bennett2013}. In outbursts of classical novae (typical temperature of 0.1 GK<$T$<0.4 GK), $^{26}$Al is produced by the reaction chain $^{24}$Mg($p,\gamma$)$^{25}$Al($\beta ^+$)$^{25}$Mg($p,\gamma$)$^{26}$Al, which however, could be bypassed via proton capture reaction of $^{25}$Al(p,$\gamma$)$^{26}$Si at high temperature (for example, $T$>0.27 GK\cite{Iliadis1996}), since the isomeric state $^{26}$Al$^m$ rather than ground state $^{26}$Al$^g$ is predominantly populated by the subsequent $\beta$ decay of $^{26}$Si\cite{Peplowski2009,Iliadis1996}. Thus, reliable measurement with proton capture reaction rate of $^{25}$Al is of great importance to better understand the origin of Galactic $^{26}$Al.

Direct measurement of $^{25}$Al($p,\gamma$)$^{26}$Si reaction cross section gives the most reliable information on the reaction rate, but current available beam intensity of $^{25}$Al is not sufficient to perform such measurement. Indirect measurements combined with theoretical calculations have been therefore conducted. Previous studies convinced that there are four important resonant states, namely $E_x=5676.2(3)$ keV, $E_x=5890.1(3)$ keV, $E_x=5929.4(8)$ keV and $E_x=5945.9(40)$ keV, which lie within 500 keV above the proton threshold of $^{26}$Si (5514.0(1) keV)\cite{datasheet2016,AME2016}, could contribute to the $^{25}$Al($p,\gamma$)$^{26}$Si reaction rate in explosive hydrogen burning environments of classical novae\cite{Iliadis1996,Wrede2009,Parpottas2004,Peplowski2009,Bardayan2006,Doherty2015,Komatsubara2014,Chipps2016}. In fact, crucial resonance information including decay properties of the $E_x=5929.4(8)$ keV state and spin parity assignment of $E_x=5945.9(40)$ keV state are still under debate. Among the four resonances, $E_x=$5929.4(8) keV $J^{\pi}=3^+$ state, which can be populated by the $\beta$ decay of $^{26}$P, has been extensively studied\cite{Parpottas2004,Peplowski2009,Bennett2013,Loureiro2016}, as this state was expected to dominate the total reaction rate at high temperature environments\cite{Iliadis1996}. In 2013, Bennett \textit{et al.}\cite{Bennett2013} observed the $\beta$-delayed $\gamma$ ray of 1742 keV emitted from the $E_x=$5929.4(8) keV $J^{\pi}=3^+$ state with $\beta\gamma$ intensity of $ \left[0.18 \pm 0.05\left(stat. \right) \pm 0.04\left(lit. \right)\right]\%  $. Based on experimentally determined absolute $\beta p$ intensity of $ I_p $=17.96(90)\% by Thomas \textit{et al.}\cite{Thomas2004} and proton-decay partial width of $\Gamma_p=2.9(10)$ eV by Peplowski \textit{et al.}\cite{Peplowski2009}, the partial width of gamma-decay was derived to be $\Gamma_\gamma=40\pm11(stat.)^{+19}_{-18}(lit.)$ meV from the relation of $I_p/I_{\gamma}=\Gamma_p/\Gamma_\gamma$. However, a recent study\cite{Janiak2017}, in which background-free proton spectrum was acquired using optical time projection chamber (OTPC), reported an inconsistent value of $ I_p $=$10.4(9)\sim13.8(10)\%$ for $\beta p$ intensity. Present work remeasured $\beta p$ and $\beta \gamma$ intensities with high accuracy, particularly the simultaneous measurement of $\beta p$ and $\beta \gamma$ would reduce the uncertainties caused by different experimental setups.

Besides, shell model calculations were also performed in this work to study the decay information of other resonant states. For $E_x=5945.9(40)$ keV state, the spin parity was reported to be $0^+$ based on the comparison between measured differential cross sections and Hauser-Feshbach calculated cross sections in the measurement of $^{24}$Mg($^3$He,n) reaction\cite{Parpottas2004}. Later in Refs.\cite{Komatsubara2014,Doherty2015}, another state at $E_x=5890.1(3)$ keV was unambiguously identified with the spin parity of $0^+$ by $\gamma$-$\gamma$ angular correlation measurements. However, the existence of two $0^+$ states within this resonance energy region was not supported by neither shell model calculations nor mirror symmetry analysis. In a recent compilation\cite{Chipps2016}, Chipps gave a detailed discussion for this puzzle and pointed out that $0^+$ assignment for both states was possible if one is due to particles being excited into a different shell. This speculation inspired us to perform shell model calculations in USD interaction taking into account the cross shell excitations to explore the spin parity assignment and decay properties of $E_x=5945.9(40)$ keV state. Moreover, partial widths of proton- and gamma-decay of other resonances at $E_x=5676.2(3)$ keV and $E_x=5890.1(3)$ keV were also calculated consistently in the same framework.

Combining the experimentally determined decay information of the $E_x=5929.4(8)$ keV $J^{\pi}=3^+$ state and the theoretical calculated properties of other resonances based on shell model with weakly bound effects, proton capture reaction rate of $^{25}$Al(p,$\gamma$)$^{26}$Si was investigated and compared with previous studies.

\section{Experimental Techniques}\label{section2}
\begin{figure}[htbp]
	\includegraphics[width=0.90\linewidth]{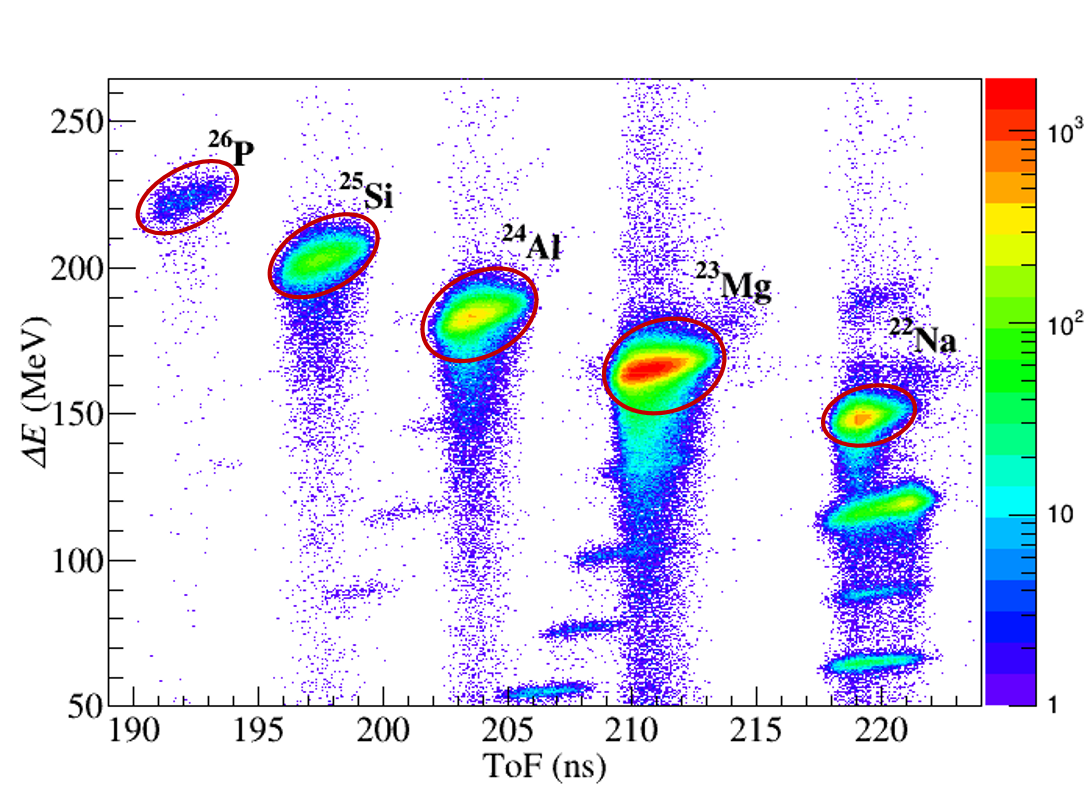}
	\caption{\label{fig:DE1TOF} Particle identification plot of ToF-$\Delta E$ obtained from the plastic scintillators and one of the silicon detectors ($\Delta E1$). Heavy ions are marked with red circles and the corresponding isotope symbols.}
\end{figure}

The experiment was carried out at the Heavy Ion Reaction Facility in Lanzhou (HIRFL)\cite{Zhan2008533c}. A 80.6 MeV/nucleon $^{32}$S$^{16+}$ primary beam was produced by the K69 Sector Focus Cyclotron and the K450 Separate Sector Cyclotron and then impinged upon a 1581 $\mu$m thick $^9$Be target to produce the secondary radioactive ions which were in-flight separated and purified by the Radioactive Ion Beam Line in Lanzhou (RIBLL1)\cite{SUN2003496}. $^{26}$P heavy ions were identified by the time-of-flight (T1 and T2 from two plastic scintillators) and energy loss ($\Delta E$1 and $\Delta E$2 from two silicon detectors) method event-by-event under a certain magnet rigidity. The two-dimensional spectrum for particle identification is presented in Fig.~\ref{fig:DE1TOF}. During the total beam time of about 95.3 hours, an average intensity of about 1 particle per second of $^{26}$P ions were delivered to the detection system.

A detection system, including three double-sided silicon strip detectors (DSSDs), five Clover-type high-purity Germanium (HPGe) detectors and five quadrant silicon detectors (QSDs), was placed at the last focal plane (T2) of RIBLL1 for the heavy ion implantation and the measurement of the subsequent decay. The basic techniques of the detection system were described in Refs.\cite{SUN20151,Sun2019}. Charged particles emitted in the decay of $^{26}$P, such as protons and $\beta$ particles, were measured by the time and position correlations between implantation and decay signals\cite{Sun2015CPL,SUN20151,Sun2019,Xu2017,Sun2017,WangK2018,WangYT20181,Xu2018} with three DSSDs of different thicknesses (DSSD1 of 142 $\mu$m, DSSD2 of 40 $\mu$m and DSSD3 of 304 $\mu$m). The thinnest silicon detector, DSSD2, which was installed between DSSD1 and DSSD3, was mainly used to detect low-energy protons for reducing the proton peak shifts due to the $\beta$-summing effect\cite{Gorres1992,TRINDER1997}. The thicker ones, DSSD1 and DSSD3, were employed for the measurement of high-energy protons and $\beta$ particles. At the upstream of DSSD array, two of the QSDs were applied to detect the energy loss of the heavy ions for particle identification. At the downstream of DSSD array, another three QSDs were installed to detect $\beta$ particles and light contaminations in the beam, such as $^1$H, $^3$He and $^4$He, for background reduction. Around the DSSD array, five HPGe detectors were used to measure the $\gamma$ rays during the $\beta$ decay of $^{26}$P. In addition to the detectors mentioned, nine movable Aluminum degraders with different thicknesses were assembled to reduce the beam energy and then enabled most of the $^{26}$P heavy ions could be stopped by the DSSDs. Totally, about $3.0\times10^5$ $^{26}$P ions were implanted into the DSSD array with proportions of 2.1\%, 45.5\%, 52.4\% in DSSD1, DSSD2 and DSSD3, respectively.

\section{Analysis and Results}\label{section3}
\subsection{Half-life of $^{26}$P}

\begin{figure}[htbp]
	\includegraphics[width=1.0\linewidth]{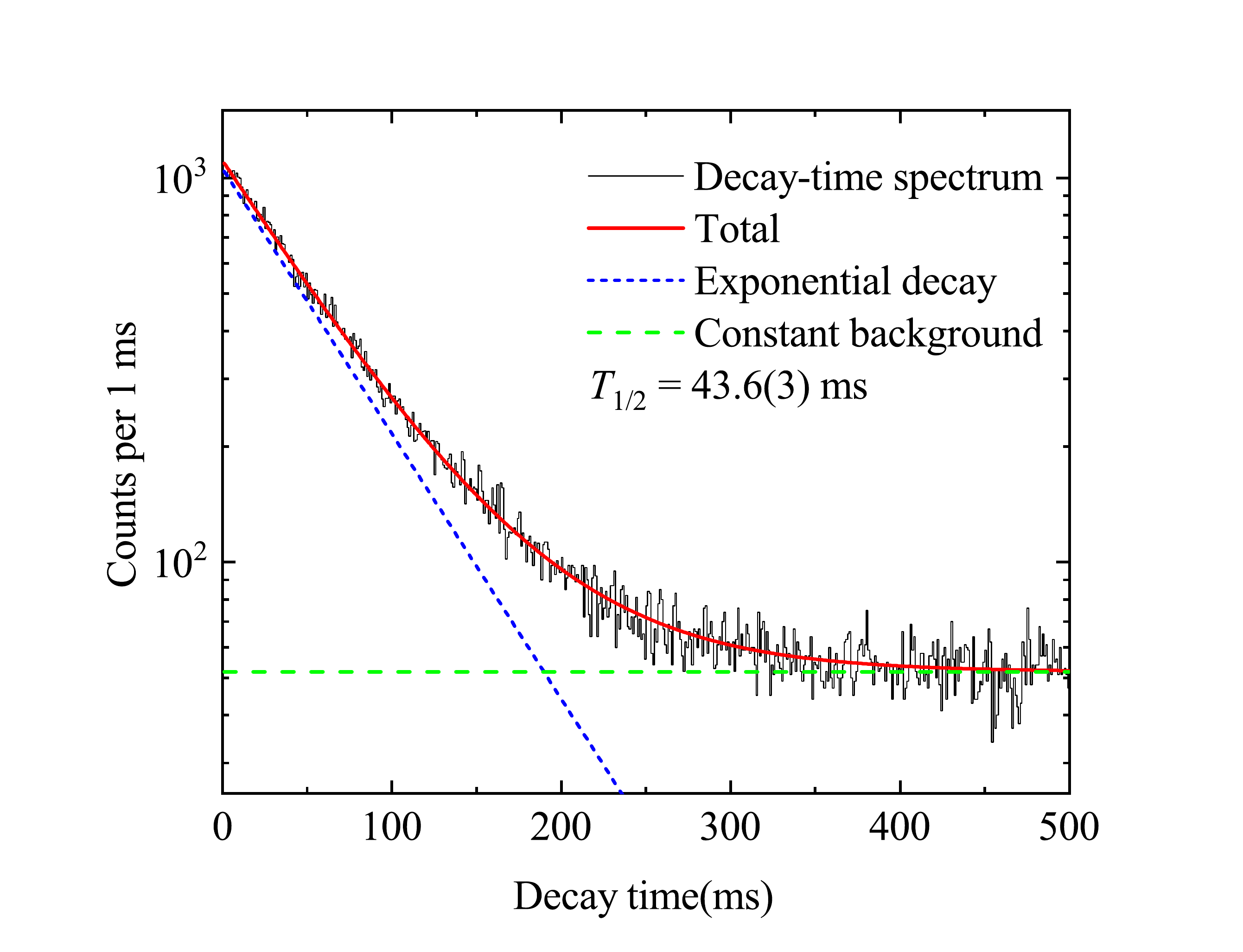}
	\caption{\label{fig:halflife} Decay-time spectrum of $^{26}$P fitted with an exponential decay component (blue short-dashed line) and a constant background component (green long-dashed line).}
\end{figure}

Figure~\ref{fig:halflife} shows the cumulative decay-time spectrum of $^{26}$P from the DSSDs. The time differences between the implantation of $^{26}$P into one DSSD and the subsequent decay events in the same pixel are fitted with an exponential decay component and a constant background component to determine the half-life of $^{26}$P. A time window of 500 ms is set here for the estimation of background which is mainly caused by randomly correlated events. Anti-coincidence from the QSDs is also applied to reduce the background from light nuclei in the beam. In present work, half-life of $^{26}$P is measured to be $T_{1/2}$=43.6(3) ms, which is in good agreement with literature value of $43.7(6)$ ms given by Thomas \textit{et al}\cite{Thomas2004}. The uncertainty of our result is directly derived from the fitting of decay-time spectrum.

\subsection{$\beta p$ and $\beta \gamma$ intensities}\label{section4}
\begin{figure}[htbp]
	\includegraphics[width=1.0\linewidth]{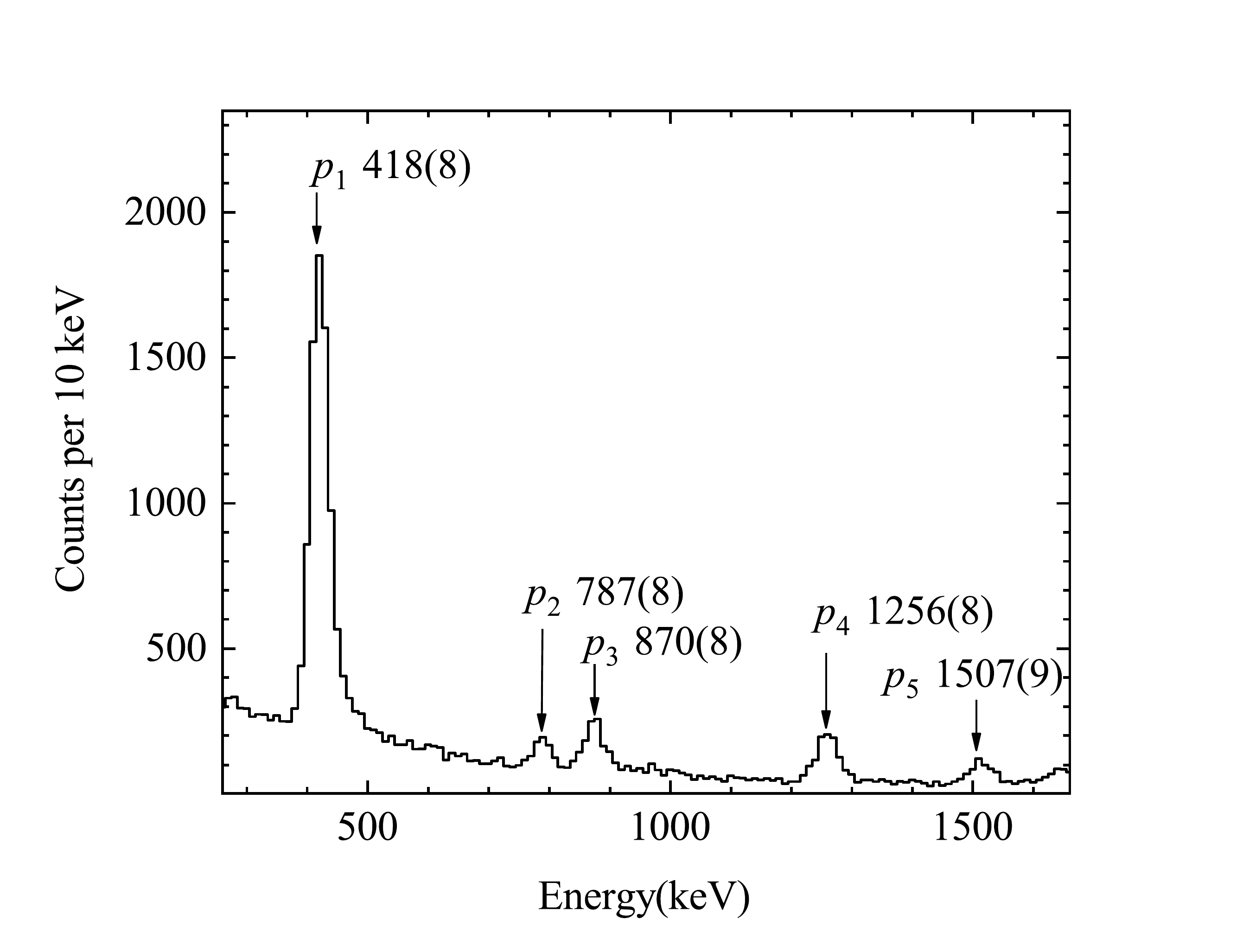}
	\caption{\label{fig:proton} Low energy part of the cumulative $\beta$-delayed charged particle spectrum from the $\beta$ decay of $^{26}$P. Proton peaks are labeled as $p_1$ to $p_5$ with the corresponding center-of-mass energy. $\beta$ pile-up effects are strongly suppressed by the anti-coincidence of $\beta$ signals from QSD1.}
\end{figure}

An inner-source method with the well-studied $\beta$-delayed protons from the decay of $^{25}$Si\cite{Thomas2004} was used for the energy calibration and detection efficiency calibration of DSSD array. The adopted energy and corresponding absolute intensity of proton peaks are 401(1) keV [4.75(32\%)], 943(2) keV [1.63(20)\%], 1804(8) keV [0.58(13)\%], 1917(2) keV [2.24(21)\%], 2162(4) keV [1.73(22)\%], 2307(4) keV [1.57(21)\%], 3463(3) keV [2.68(26)\%], 4252(2) keV [9.54(66)\%], and 5624(3) keV [2.39(20)\%]\cite{Thomas2004}. Figure \ref{fig:proton} displays the low energy part of the cumulative charged-particle spectrum, as the peak labeled as $p_1$ represents the $\beta$-delayed protons emitted from the $E_x=5929.4(8)$ keV $3^+$ state of $^{26}$Si to the ground state of $^{25}$Al, which can be used to calculate the resonance energy and the partial width. The $\beta$ particles in QSD1 is applied as anti-coincidence to suppress the $\beta$ pile-up effects, thus improving the energy resolution of the proton peaks. The energy of the first proton peak is determined to be $p_1$=418(8) keV, which is consistent with  literature values of 412(2) keV\cite{Thomas2004}, 426(30) keV\cite{Janiak2017} and the derived value of 415.4(8) keV from the databases\cite{AME2016,datasheet2016}. The uncertainty of proton peak energy includes the statistical uncertainty of 0.33 keV from the fitting of spectrum and the systematical uncertainty of 7.25 keV that is attributed from calibration. In present work, the absolute intensity of $p_1$ is measured to be $I_{p1}=11.1(12)\%$, which is in good agreement with a recent study of $10.4(9)\sim13.8(10)\%$\cite{Janiak2017}, but inconsistent with the value of $17.96(90)\%$ in Ref.\cite{Thomas2004} as shown in the second column of Table \ref{tab:intensity}. The uncertainty here is mainly caused by the statistical uncertainty of 0.15\% and the systematic uncertainties of 1.14\% from the calibration. Other proton peaks at $p_2$=787(8) keV, $p_3$=870(8) keV, $p_4$=1256(8) keV and $p_5$=1507(9) keV are all observed clearly with corresponding absolute intensities of 0.74(17)\%, 1.44(30)\%, 1.45(21)\% and 0.80(18)\%, respectively.
\renewcommand\arraystretch{1.5}
\begin{table}
	\newcommand{\tabincell}[2]{\begin{tabular}{@{}#1@{}}#2\end{tabular}}
	\caption{\label{tab:intensity}$\beta p$ and $\beta\gamma$ intensities of $E_x=5929.4(8)$ keV $3^+$ state of $^{26}$Si from this work and previous studies.}
	\begin{ruledtabular}
		\begin{tabular}{ccc}
			Reference&$I_{p1}$(\%)[418(8) keV]&$I_{\gamma1}$(\%)[1742(2) keV]\\ \hline
			This work &\textbf{11.1(12)} &  \textbf{0.59(44)} \\
			Thomas\cite{Thomas2004}& 17.96(90) & \\
			Janiak\cite{Janiak2017}& $10.4(9)\sim13.8(10)$&  \\
			Bennett\cite{Bennett2013}& & \tabincell{c}{$[0.18 \pm 0.05\left(stat. \right)$ \\ $\pm 0.04\left(lit. \right)] $}  \\
			P{\'e}rez-Loureiro\cite{Loureiro2016}& &0.15(5)  \\
		\end{tabular}
	\end{ruledtabular}
\end{table}
\begin{figure}[htbp]
	\includegraphics[width=1.0\linewidth]{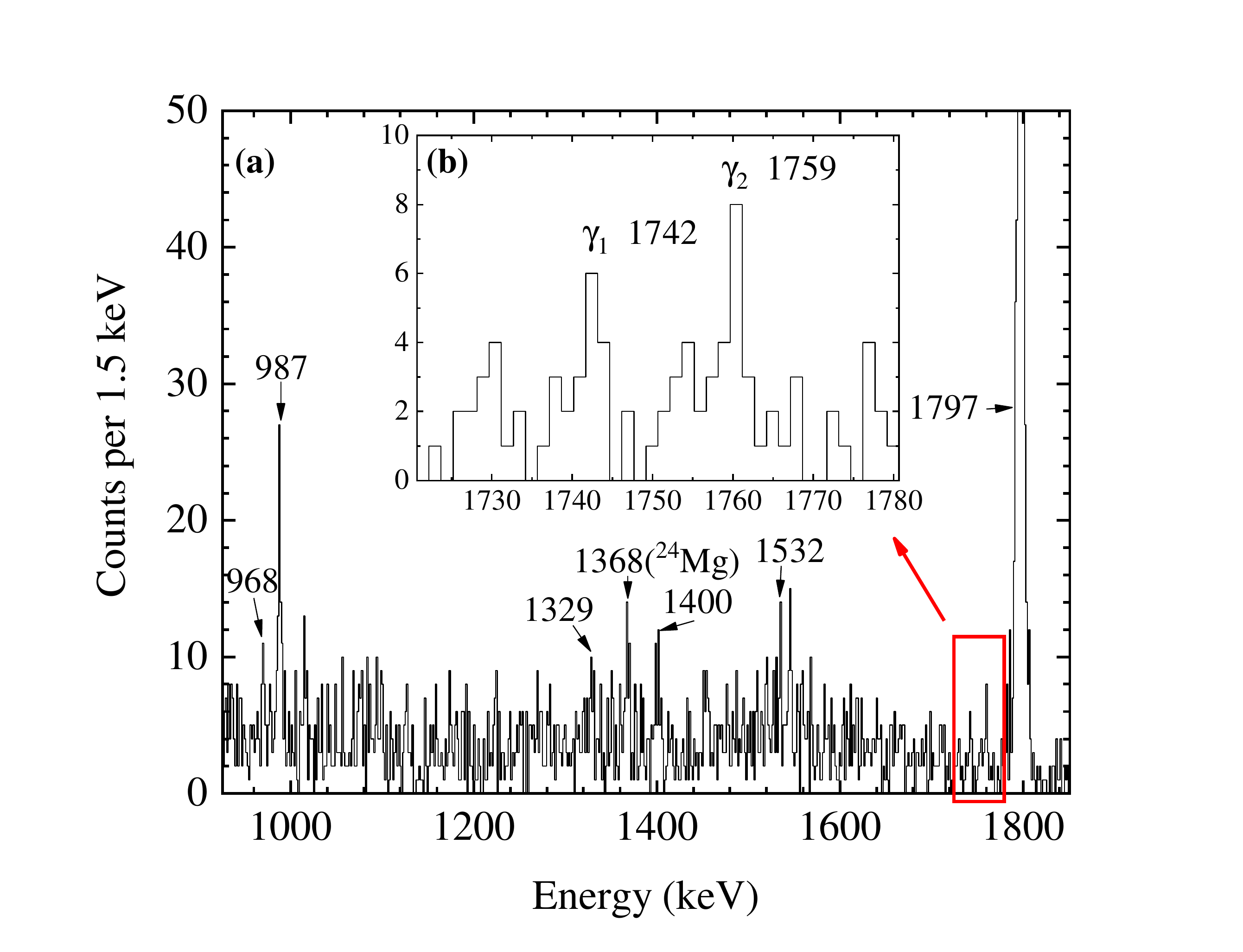}
	\caption{\label{fig:wide1} (a) The cumulative $\gamma$-ray spectrum measured by HPGe detectors in coincidence with $\beta$ particles detected by DSSD3. (b) $\gamma$-ray spectrum from 1720 keV to 1780 keV.}
\end{figure}

For HPGe detectors, the energy calibration and intrinsic detection efficiency calibration were performed with four standard sources $^{152}$Eu, $^{133}$Ba, $^{60}$Co, and $^{137}$Cs.  As $^{22}$Al was also studied with the same detection configurations in the experiment, the absolute detection efficiency could be deduced by the $\beta$-delayed $\gamma$-ray transitions with known energies and absolute intensities of 988 keV [5.7(3)\%], 1796 keV [58(3)\%] from $^{26}$P\cite{Loureiro2016}, 452 keV [18.4(42)\%], 493 keV [15.3(34)\%], 945 keV [10.4(23)\%], 1612 keV [15.2(32)\%] from $^{25}$Si\cite{Thomas2004} and 1248.5 keV [38.2(69)\%], 1985.6 keV [31.1(54)\%]), 2062.3 keV [34.1(58)\%] from $^{22}$Al\cite{Achouri2006}. Figure \ref{fig:wide1}(a) shows the low background $\gamma$-ray spectrum measured by the HPGe detectors in coincidence with the $\beta$ signals of $^{26}$P in DSSD3. Strong $\gamma$-ray transitions at 987 keV (2$^{+}_2$ $\rightarrow$2$^{+}_1$) and 1797 keV (2$^{+}_1$ $\rightarrow$$0^+_1$(g.s.)) and also the $\gamma$-ray transitions with weak intensities, such as 968 keV (3$^{+}_1$ $\rightarrow$$2^+_2$), 1400 keV (3$^{+}_2$ $\rightarrow$$2^+_2$), 1329 keV (4$^{+}_4$ $\rightarrow$$3^+_2$) and 1532 keV (4$^{+}_3$ $\rightarrow$$3^+_1$), can be observed clearly in the spectrum. The zoomed in spectrum with energy region from 1720 keV to 1780 keV is shown in Fig.~\ref{fig:wide1}(b). $\gamma_{1}$ is the transition from 3$^{+}_3$ ($E_x=5929.4(8)$ keV) state to $3^+_2$ state and the energy is measured to be $\gamma_{1}$=1742(2) keV in this work. Here the uncertainty is the quadratic sum of statistical uncertainty of 1.1 keV from the peak and systematical uncertainties of 1.5 keV from the calibration. A constant flat background is estimated for the spectrum and the intensity of $\gamma_{1}$ is determined to be $I_{\gamma{_1}}=0.59(44)\% $, where the uncertainty is the quadratic sum of statistical uncertainty of 0.43\% from the counts and systematic uncertainties of 0.09\% from the calibration. Another peak labeled as $\gamma_{2}$ represents the transition from 4$^{+}_4$ state to $3^+_1$ state, and the energy and intensity are determined to be $\gamma_{2}$=1759(2) keV and $I_{\gamma{_2}}=0.86(51)\% $, respectively. Present result is consistent with previous studies of $I_{\gamma{_1}}=\left[0.18 \pm 0.05\left(stat \right) \pm 0.04\left(lit. \right)\right]\% $ in Ref.\cite{Bennett2013} and $I_{\gamma{_1}}=0.15(5)\%$ in Ref.\cite{Loureiro2016} as presented in the third column of Table \ref{tab:intensity}. The partial gamma-decay branch of the 1742(2) keV $\gamma$-ray from the $E_x=$5929.4(8) keV $J^{\pi}=3^+$ level was expected to be $71_{-19}^{+13}$\% in Refs.\cite{Wrede2009,Bennett2013}. Adopting this assumption yields a total $\beta\gamma$ intensity of $I_{\gamma}=0.83_{-0.64}^{+0.66}$\% for all primary gamma rays from this state. Together with the $\beta p$ intensity of $I_{p1}=11.1(12)\%$ and the experimentally determined proton-decay partial width of $\Gamma_p=2.9(10)$ eV in Ref.\cite{Peplowski2009}, the gamma-decay partial width is derived to be $\Gamma_\gamma =0.22^{+0.19}_{-0.18}$ eV using the relation $I_{p}/I_{\gamma}=\Gamma_p/\Gamma_\gamma$.

\subsection{Shell model calculation}\label{section5}
\renewcommand\arraystretch{1.5}
\begin{table*}
	\caption{\label{tab:table1}Present  $^{25}$Al$(p,\gamma)^{26}$Si resonance parameters.}
	\begin{ruledtabular}
		\begin{tabular}{ccccccc}
			$J^{\pi}$&$E_x$(keV)\footnote{Adopted $E_x$ from the database\cite{datasheet2016}.}&$E_r=E_x-Q_{p \gamma}$(keV)\footnote{$Q_{p \gamma}=5514.0(1)$ keV derived from AME2016\cite{AME2016}.}&$E_r$(keV) adopted&$\Gamma_\gamma$(eV)& $C^2S_p$ &$\Gamma_p$(eV)\\ \hline
			$1^{+}$&5676.2(3)&162.2(3) &162.2(3)&$1.17\times10^{-1}$& 7.00$\times 10^{-3}$ &$1.25$$\times10^{-8}$\\
			$0{^+}$&5890.1(3)&376.1(3)&376.1(3)&$7.71\times10^{-3}$& $3.86\times10^{-2}$ &$3.86\times10^{-3}$\\
			$3{^+}$&5929.4(8)&415.4(8)&\textbf{418(8)}\footnote{This work.}&$\mathbf{2.16^{+1.89}_{-1.84}\times10^{-1}}$& &2.9(10)\footnote{Adopted values in Ref.\cite{Peplowski2009}.}\\
			($4{^+}$) &5945.9(40)&431.9(40)&431.9(40)&$2.50\times10^{-2}$& $1.96\times10^{-2}$ &$7.80\times10^{-3}$\\
			
		\end{tabular}
	\end{ruledtabular}
\end{table*}

The spin-parity of 5945.9(40) keV state was reported to be $0^+$ by Parpottas \textit{et al.}\cite{Parpottas2004} which was concluded from the comparison of experimental cross sections with Hauser-Feshbach calculations. But it was under debate as only one $0^+$ state was expected in this resonant energy region based on the mirror nuclei analysis and shell model calculations\cite{Doherty2015,Komatsubara2014,Chipps2016}. In Ref.\cite{Chipps2016}, Chipps suggested that a $0^+$ assignment for both states is possible if one is due to cross shell excitations. Therefore, present work performed shell model calculations in both $sd$ region and cross shell excitations to investigate the spin-parity assignment of $E_x=5945.9(40)$ keV state. 

USD family including USD\cite{Wildenthal1984}, USDA and USDB \cite{Brown2006,Brown1988} is successful Hamiltonians in $sd$ region. But they are developed for the structure of neutron-rich nuclei. If proton-rich nuclei are considered, such as $^{26}$Si and $^{25}$Al, the weakly bound effect of the proton $1s_{1/2}$ orbit should be included \cite{Yuan2014}. USD*, USDA* and USDB* Hamiltonians which incorporate such effect reasonably reproduce the mirror energy differences in $sd$ region \cite{Yuan2014}. Recent observations on $\beta$-decays of $^{22}$Si and $^{27}$S show that the weakly bound effect is important to explain the decay properties and the levels of corresponding daughter nuclei\cite{Xu2017,Sun2019}. 
In our calculation, all three Hamiltonians in $sd$ region, USD*, USDA* and USDB*, give similar results for the structure of $^{26}$Si, which also reproduce the experimental energy and spin-parity of present considered states\cite{datasheet2016}, 5676.2(3) keV ($1^{+}_{1}$), 5890.1(3) keV ($0^{+}_{4}$), and 5929.4(8) keV ($3^{+}_{3}$). However, other than the $0^{+}_{4}$ state, all three Hamiltonians predict that the $0^{+}_5$ state locates at $E_x$>7.9 MeV. Further shell model calculations, in which the $p$ to $sd$ and $sd$ to $pf$ cross shell excitations through the $psd$ Hamiltonian YSOX \cite{Yuan2012} and $sdpf$ Hamiltonian sdpf-m \cite{Utsuno1999} are considered, are performed to attempt to explain the possibility of low-lying $0^+_5$ state. But the results show that the cross shell excitation could not reduce the excitations energy of $0^{+}_{5}$ state to be lower than $E_x$=7.7 MeV. Indeed, all three Hamiltonians predict a $4^+$ state around $E_x$=5.8 MeV which is consistent with the mirror nuclei analysis\cite{Richter2011}. Therefore, our calculations refute the existence of another $0^+$ state within the interested resonance energy region, but favor a $4^{+}$ assignment for 5945.9(40) keV state. 

Table~\ref{tab:table1} shows the resonance parameters adopted in present work. Except for the decay properties of $E_x$=5929.4(8) keV state, proton- and gamma-decay partial widths of other resonant states are all calculated values with above-mentioned shell model in $sd$ region. Combining the observed decay energies and shell model B(E2) and B(M1) values, the gamma-decay partial widths of 5676.2(3) keV, 5890.1(3) keV and 5945.9(40) keV states are calculated to be 1.17$\times$ 10$^{-1}$ eV, 7.71$\times$ 10$^{-3}$ eV and 2.50$\times$ 10$^{-2}$ eV, respectively, as shown in the fifth column of Table \ref{tab:table1}. To calculate the proton-decay partial width, equation (\ref{equation:protonwidth})\cite{Iliadis2008} can be used
\begin{eqnarray}
\Gamma_p=C^2S_p\Gamma_{sp}\label{equation:protonwidth}
\end{eqnarray}
where $C^2S_p$ is the single-particle spectroscopic factor calculated by the shell model in $sd$ region mentioned above, and $\Gamma_{sp}$ is the single-particle partial width. The last column of Table \ref{tab:table1} shows the proton-decay partial widths adopted in this work.

\subsection{Astrophysical reaction rate}

The total reaction rate for $^{25}$Al$(p,\gamma)^{26}$Si can be expressed as the sum of all resonant and nonresonant capture contributions. For the resonant part, the reaction rate can be calculated by the well-known narrow resonance formalism\cite{Rolfs1988,He2017},
\begin{eqnarray}
N_A\left\langle \sigma \nu\right\rangle_r=1.5394\times10^{11}\left(\mu T_9 \right)^{-\frac{3}{2}}(\omega\gamma)\exp\left(-\frac{11.605E_r}{T_9} \right)\nonumber\\
\label{equation:one}
\end{eqnarray}
where $\mu=A_T/(1+A_T)$ is the reduced mass in atomic mass units, and $A_T$=25 is the mass number of $^{25}$Al. $T_9$ is the temperature in units of GK and $E_r$ is the resonance energy in MeV. $\omega\gamma$ is the resonance strength in MeV, which is described as:  
\begin{eqnarray}
\omega\gamma={\frac{2J_r+1}{(2J_p+1)(2J_T+1)}}{\frac{\Gamma_p\Gamma_\gamma}{\Gamma_{\mathrm{tot}}}}
\label{equation:two}
\end{eqnarray}
where $J_r$ is the spin of resonance, $J_p=\frac{1}{2}$ is the spin of proton and $J_T$=$\frac{5}{2}$ is the spin of the ground state of $^{25}$Al. $\Gamma_p$ and $\Gamma_\gamma$ are the proton and $\gamma$-ray partial widths of the resonance, respectively, and $\Gamma_{\mathrm{tot}}$ is the total width defined as $\Gamma_{\mathrm{tot}}$=$\Gamma_p+\Gamma_\gamma$.

The resonance energy $E_r$ plays an important role in the calculation of resonant capture reaction rate as it is exponentially related in the narrow resonance formalism (see equation (\ref{equation:one})). In Table \ref{tab:table1}, the resonance energy of $E_x=$5929.4(8) keV $J^{\pi}=3^+$ level is experimentally determined with $E_r$=418(8) keV in this work (proton peak $p_1$ in Fig.~\ref{fig:proton}), while other resonance energies are all derived values by relation $E_r=E_x-Q_{p \gamma}$. As the reaction Q value can be derived precisely from the database AME2016\cite{AME2016}, the uncertainties of resonance energy are dominated by the errors in excitation energy $E_x$. In previous studies, different excitation energies were measured with various experimental techniques. Here we adopt the values in the latest database\cite{datasheet2016} which are evaluated from the previous measurements.         

The nonresonant part of $^{25}$Al$(p,\gamma)^{26}$Si reaction rate can be estimated using the expression\cite{He2017,Rolfs1988,Herndl1995}:
\begin{eqnarray}
N_A\left\langle \sigma \nu\right\rangle_{dc}=7.8327\times10^{9}\left(\frac{Z_T}{\mu T_9^2} \right)^\frac{1}{3}S_{eff}  \nonumber\\
\times \exp\left[-4.2487\left(\frac{Z_T^2\mu}{T_9} \right)^\frac{1}{3}\right] 
\label{equation:three}
\end{eqnarray}
where $\mu$ is the reduced mass in atomic mass units and $Z_T$=13 is the atomic number of $^{25}$Al. $S_{eff}$ is the effective astrophysical $S$ factor that can be expressed by
\begin{eqnarray}
S_{eff}\approx S(0)\left[1+0.09807(\frac{T_9}{Z_T^2\mu})^{1/3}\right]
\end{eqnarray}
where $S(0)$ is the $S$ factor at zero energy\cite{He2017,Fowler1967,Rolfs1988}. In previous works, both Iliadis \textit{et al}\cite{Iliadis1996} and Matic \textit{et al}\cite{Matic2010} calculated the $S$ factor with the value of 27 keV-b and 28 keV-b, respectively. As the $Q$ value adopted in Matic's calculation was much closer to the derived value of $Q_{p\gamma}$=5514.0(1) keV by AME2016\cite{AME2016}, we adopt $S(0)$=28 keV-b in Matic's work for the calculation of nonresonant capture reaction rate. A 30\% uncertainty of the nonresonant capture reaction rate is used here to estimate the upper and lower limit following Ref.\cite{Iliadis2001}.

\begin{figure}[htbp] 
	\includegraphics[width=1.0\linewidth]{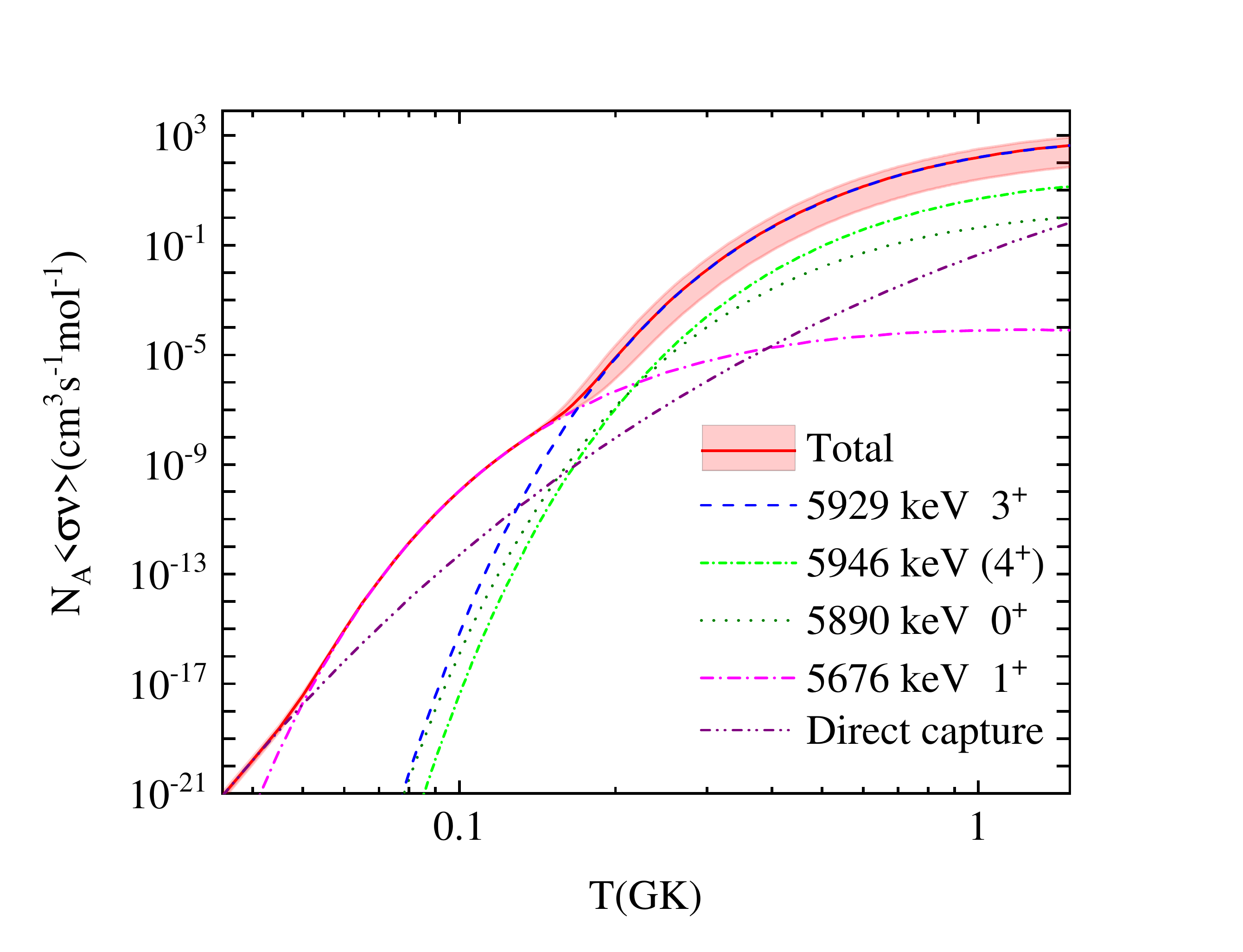}
	\caption{\label{fig:reactionrate} Proton capture reaction rates of $^{25}$Al$(p,\gamma)^{26}$Si. The red shadow area is the upper and lower limit of total reaction rate. Nonresonant and resonant reaction rates are presented in different colored lines. }
\end{figure}

Figure \ref{fig:reactionrate} presents the $^{25}$Al$(p,\gamma)^{26}$Si reaction rates as a function of the stellar temperature $T$ in units of GK. The shadow area coloured in red displays the upper and lower limit of total reaction rate. The nonresonant part makes the largest contribution to the total reaction rate when $T$<0.05 GK. Then the $E_x=5676.2(3)$ keV $J^{\pi}=1^+$ resonance becomes the main component of the total rate until $T\approx$0.16 GK. In temperature range of $T$>0.16 GK, the total reaction rate is dominated by the $E_x=5929.4(8)$ keV $J^{\pi}=3^+$ resonance. Here, the uncertainty is mainly caused by the uncertainty in gamma-decay partial width which is the result of low statistics of $\gamma_{1}$ at 1742(2) keV. The recently confirmed $E_x=5890.1(3)$ keV $J^{\pi}=0^+$ resonance makes non-negligible contribution to the total rate as temperature increases. When $T$>0.22 GK, the $E_x=5945.9(40)$ keV state, of which the spin parity is assigned to be $4^+$ in the discussion above, makes the secondary contribution to the total reaction rate. 
\begin{figure}[htbp] 
	\includegraphics[width=1.0\linewidth]{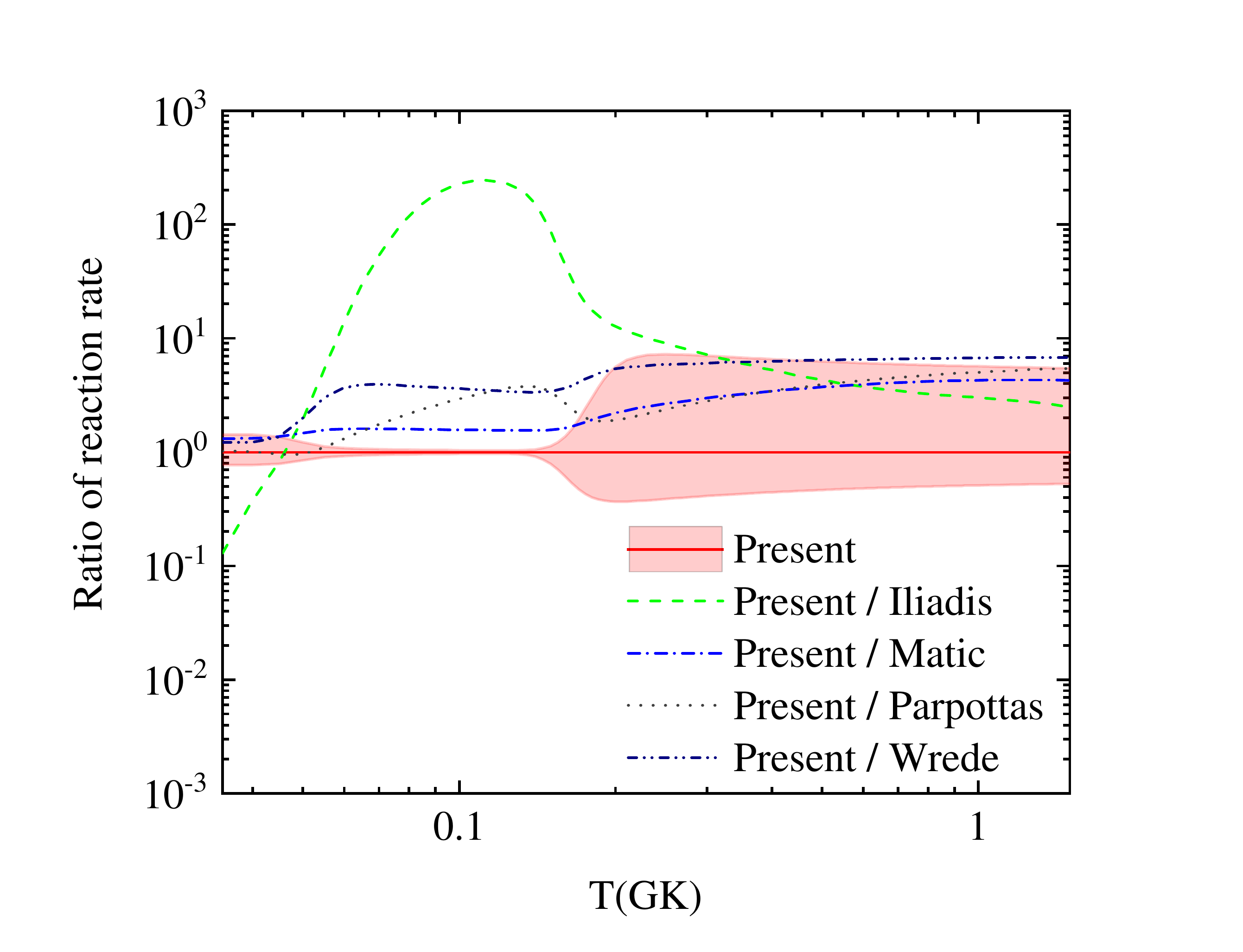}
	\caption{\label{fig:rateratio}  Ratios of present total $^{25}$Al$(p,\gamma)^{26}$Si reaction rate to the literature values\cite{Iliadis1996,Parpottas2004,Matic2010,Wrede2009} in the JINA REACLIB database\cite{Cyburt2010}. The red shadow area shows the upper and lower limit in this work. }
\end{figure}

Figure \ref{fig:rateratio} shows the ratios of present total reaction rate to the recommended literature values in the JINA REACLIB database\cite{Cyburt2010}. The shadow parts are marked as the upper and lower limit of the total reaction rate from present work. At high temperature environments where the $E_x$=5929.4(8) keV $J^{\pi}=3^+$ resonance dominates the $^{25}$Al$(p,\gamma)^{26}$Si reaction as shown in Fig. \ref{fig:reactionrate}, the total reaction rate in this work is consistent within uncertainties with the literature values while the absolute value is a factor of 3 to 6 times larger when $T$>0.20 GK. At temperature range of 0.05 GK<$T$<0.16 GK, our result is slightly larger than previous calculations as a relatively larger resonance strength of $\omega\gamma$=3.12$\times 10^{-9}$ eV is adopted for $E_x=5676.2(3)$ keV $J^{\pi}=1^+$ resonance here. At temperatures below 0.05 GK, present work shows a consistent result with previous ones. It should be noticed that the present total reaction rate could not match to the result in Iliadis' estimation\cite{Iliadis1996} at temperatures where $E_x=5676.2(3)$ keV $J^{\pi}=1^+$ resonance dominates the total reaction rate, as a much more accurate resonance energy $E_r$=162.2(3) keV from the latest databases\cite{datasheet2016,AME2016} being adopted here rather than the $E_r$=44(28) keV in Iliadis' work. The impact of present new determined $^{25}$Al$(p,\gamma)^{26}$Si reaction rate on the synthesis of $^{26}$Al in classical novae and also type-I X-ray bursts (XRBs) worths further discussion in the future.

\section{Conclusions}

In present work, we performed the first simultaneous measurement of the $\beta$-delayed protons at 418(8) keV and $\gamma$ rays at 1742(2) keV emitted from the astrophysically important $E_x$=5929.4(8) keV $J^{\pi}=3^+$ state of $^{26}$Si which dominants the proton capture reaction rate of $^{25}$Al$(p,\gamma)^{26}$Si and further influences the nucleosynthesis of Galactic $^{26}$Al. The corresponding $\beta p$ and $\beta \gamma$ intensities were measured to be $I_{p1}=11.1(12)\%$ and $I_{\gamma1}=0.59(44)\%$, respectively, with a detector system consisting of silicon array and five clover-type HPGe detectors. This simultaneous measurement could reduce the uncertainties caused by the differences in experimental setups, thus providing more reliable decay information of $E_x$=5929.4(8) keV state. Moreover, shell models with weakly bound effects in $sd$ shell region were used to investigate the resonances of $^{26}$Si and successfully reproduced the energy level and spin-parity of experimental determined states at 5676.2(3) keV ($J^\pi=1^+$), 5890.1(3) keV ($J^\pi=0^+$) and 5929.4(8) keV ($J^\pi=3^+$). On the other hand, shell model calculations with three Hamiltonians, USD*, USDA* and USDB*, in both $sd$ shell region and cross shell excitations could not reproduce a $0^{+}_{5}$ state with the excitation energy lower than $E_x$=7.7 MeV. Indeed a $4^{+}$ state at around $E_x$=5.8 MeV was predicted, suggesting a $4^{+}$ spin-parity assignment for $E_x=5945.9(40)$ keV state. By combining experimental results and theoretical calculations, we calculated the total reaction rate of $^{25}$Al$(p,\gamma)^{26}$Si in explosive hydrogen burning environments. Compared with literature values in JINA REACLIB database, our result was consistent within uncertainties with the previous studies while the absolute value at $T$>0.20 GK temperatures was a factor of 3 to 6 times larger.

\begin{acknowledgments}
We wish to acknowledge the support of the HIRFL operations staff for providing good-quality beams and the effort of the RIBLL1 collaborators for performing the experiment. This work is supported by the Ministry of Science and Technology of China under the National Key R\&D Programs No. 2016YFA0400503 and No. 2018YFA0404404, and by the National Natural Science Foundation of China under Grants No. U1932206, No. U1632136, No. 11805120, No. 11805280, No. 11825504, No. 11811530071, No. 11705244, No. 11705285, No. 11775316, No. U1732145, No. 11635015, No. 11675229, No. 11505293, No. 11475263, No. 11490562, and No. U1432246, and by the Youth Innovation Promotion Association of Chinese Academy of Sciences under Grant No. 2019406, and by the China Postdoctoral Science Foundation under Grants No. 2017M621442 and No. 2017M621035, and by the Office of China Postdoctoral Council under the International Postdoctoral Exchange Fellowship Program (Talent-Dispatch Program) No. 20180068.
\end{acknowledgments}

\nocite{*}

\bibliography{P26Bib}

\providecommand{\noopsort}[1]{}\providecommand{\singleletter}[1]{#1}%
\begin{thebibliography}{60}%
\makeatletter
\providecommand \@ifxundefined [1]{%
 \@ifx{#1\undefined}
}%
\providecommand \@ifnum [1]{%
 \ifnum #1\expandafter \@firstoftwo
 \else \expandafter \@secondoftwo
 \fi
}%
\providecommand \@ifx [1]{%
 \ifx #1\expandafter \@firstoftwo
 \else \expandafter \@secondoftwo
 \fi
}%
\providecommand \natexlab [1]{#1}%
\providecommand \enquote  [1]{``#1''}%
\providecommand \bibnamefont  [1]{#1}%
\providecommand \bibfnamefont [1]{#1}%
\providecommand \citenamefont [1]{#1}%
\providecommand \href@noop [0]{\@secondoftwo}%
\providecommand \href [0]{\begingroup \@sanitize@url \@href}%
\providecommand \@href[1]{\@@startlink{#1}\@@href}%
\providecommand \@@href[1]{\endgroup#1\@@endlink}%
\providecommand \@sanitize@url [0]{\catcode `\\12\catcode `\$12\catcode
  `\&12\catcode `\#12\catcode `\^12\catcode `\_12\catcode `\%12\relax}%
\providecommand \@@startlink[1]{}%
\providecommand \@@endlink[0]{}%
\providecommand \url  [0]{\begingroup\@sanitize@url \@url }%
\providecommand \@url [1]{\endgroup\@href {#1}{\urlprefix }}%
\providecommand \urlprefix  [0]{URL }%
\providecommand \Eprint [0]{\href }%
\providecommand \doibase [0]{http://dx.doi.org/}%
\providecommand \selectlanguage [0]{\@gobble}%
\providecommand \bibinfo  [0]{\@secondoftwo}%
\providecommand \bibfield  [0]{\@secondoftwo}%
\providecommand \translation [1]{[#1]}%
\providecommand \BibitemOpen [0]{}%
\providecommand \bibitemStop [0]{}%
\providecommand \bibitemNoStop [0]{.\EOS\space}%
\providecommand \EOS [0]{\spacefactor3000\relax}%
\providecommand \BibitemShut  [1]{\csname bibitem#1\endcsname}%
\let\auto@bib@innerbib\@empty
\bibitem [{\citenamefont {Wang}\ \emph {et~al.}(2009)\citenamefont {Wang},
  \citenamefont {Lang}, \citenamefont {Diehl}, \citenamefont {Halloin},
  \citenamefont {Jean}, \citenamefont {Kn{\"o}dlseder}, \citenamefont
  {Kretschmer}, \citenamefont {Martin}, \citenamefont {Roques}, \citenamefont
  {Strong}, \citenamefont {Winkler},\ and\ \citenamefont {Zhang}}]{Wang2009}%
  \BibitemOpen
  \bibfield  {author} {\bibinfo {author} {\bibfnamefont {W.}~\bibnamefont
  {Wang}}, \bibinfo {author} {\bibfnamefont {M.~G.}\ \bibnamefont {Lang}},
  \bibinfo {author} {\bibfnamefont {R.}~\bibnamefont {Diehl}}, \bibinfo
  {author} {\bibfnamefont {H.}~\bibnamefont {Halloin}}, \bibinfo {author}
  {\bibfnamefont {P.}~\bibnamefont {Jean}}, \bibinfo {author} {\bibfnamefont
  {J.}~\bibnamefont {Kn{\"o}dlseder}}, \bibinfo {author} {\bibfnamefont
  {K.}~\bibnamefont {Kretschmer}}, \bibinfo {author} {\bibfnamefont
  {P.}~\bibnamefont {Martin}}, \bibinfo {author} {\bibfnamefont {J.~P.}\
  \bibnamefont {Roques}}, \bibinfo {author} {\bibfnamefont {A.~W.}\
  \bibnamefont {Strong}}, \bibinfo {author} {\bibfnamefont {C.}~\bibnamefont
  {Winkler}}, \ and\ \bibinfo {author} {\bibfnamefont {X.~L.}\ \bibnamefont
  {Zhang}},\ }\href {\doibase 10.1051/0004-6361/200811175} {\bibfield
  {journal} {\bibinfo  {journal} {Astron. Astrophys.}\ }\textbf {\bibinfo
  {volume} {496}},\ \bibinfo {pages} {713} (\bibinfo {year}
  {2009})}\BibitemShut {NoStop}%
\bibitem [{\citenamefont {Diehl}\ \emph {et~al.}(2006)\citenamefont {Diehl},
  \citenamefont {Halloin}, \citenamefont {Kretschmer}, \citenamefont {Lichti},
  \citenamefont {Sch{\"o}nfelder}, \citenamefont {Strong}, \citenamefont {von
  Kienlin}, \citenamefont {Wang}, \citenamefont {Jean}, \citenamefont
  {Kn{\"o}dlseder}, \citenamefont {Roques}, \citenamefont {Weidenspointner},
  \citenamefont {Schanne}, \citenamefont {Hartmann}, \citenamefont {Winkler},\
  and\ \citenamefont {Wunderer}}]{Diehl2006}%
  \BibitemOpen
  \bibfield  {author} {\bibinfo {author} {\bibfnamefont {R.}~\bibnamefont
  {Diehl}}, \bibinfo {author} {\bibfnamefont {H.}~\bibnamefont {Halloin}},
  \bibinfo {author} {\bibfnamefont {K.}~\bibnamefont {Kretschmer}}, \bibinfo
  {author} {\bibfnamefont {G.~G.}\ \bibnamefont {Lichti}}, \bibinfo {author}
  {\bibfnamefont {V.}~\bibnamefont {Sch{\"o}nfelder}}, \bibinfo {author}
  {\bibfnamefont {A.~W.}\ \bibnamefont {Strong}}, \bibinfo {author}
  {\bibfnamefont {A.}~\bibnamefont {von Kienlin}}, \bibinfo {author}
  {\bibfnamefont {W.}~\bibnamefont {Wang}}, \bibinfo {author} {\bibfnamefont
  {P.}~\bibnamefont {Jean}}, \bibinfo {author} {\bibfnamefont {J.}~\bibnamefont
  {Kn{\"o}dlseder}}, \bibinfo {author} {\bibfnamefont {J.-P.}\ \bibnamefont
  {Roques}}, \bibinfo {author} {\bibfnamefont {G.}~\bibnamefont
  {Weidenspointner}}, \bibinfo {author} {\bibfnamefont {S.}~\bibnamefont
  {Schanne}}, \bibinfo {author} {\bibfnamefont {D.~H.}\ \bibnamefont
  {Hartmann}}, \bibinfo {author} {\bibfnamefont {C.}~\bibnamefont {Winkler}}, \
  and\ \bibinfo {author} {\bibfnamefont {C.}~\bibnamefont {Wunderer}},\ }\href
  {\doibase https://doi.org/10.1038/nature04364} {\bibfield  {journal}
  {\bibinfo  {journal} {Nature}\ }\textbf {\bibinfo {volume} {439}},\ \bibinfo
  {pages} {45} (\bibinfo {year} {2006})}\BibitemShut {NoStop}%
\bibitem [{\citenamefont {{Kn{\"o}dlseder}}\ \emph {et~al.}(1999)\citenamefont
  {{Kn{\"o}dlseder}}, \citenamefont {{Bennett}}, \citenamefont {{Bloemen}},
  \citenamefont {{Diehl}}, \citenamefont {{Hermsen}}, \citenamefont
  {{Oberlack}}, \citenamefont {{Ryan}}, \citenamefont {{Sch{\"o}nfelder}},\
  and\ \citenamefont {{von Ballmoos}}}]{Knodlseder1999}%
  \BibitemOpen
  \bibfield  {author} {\bibinfo {author} {\bibfnamefont {J.}~\bibnamefont
  {{Kn{\"o}dlseder}}}, \bibinfo {author} {\bibfnamefont {K.}~\bibnamefont
  {{Bennett}}}, \bibinfo {author} {\bibfnamefont {H.}~\bibnamefont
  {{Bloemen}}}, \bibinfo {author} {\bibfnamefont {R.}~\bibnamefont {{Diehl}}},
  \bibinfo {author} {\bibfnamefont {W.}~\bibnamefont {{Hermsen}}}, \bibinfo
  {author} {\bibfnamefont {U.}~\bibnamefont {{Oberlack}}}, \bibinfo {author}
  {\bibfnamefont {J.}~\bibnamefont {{Ryan}}}, \bibinfo {author} {\bibfnamefont
  {V.}~\bibnamefont {{Sch{\"o}nfelder}}}, \ and\ \bibinfo {author}
  {\bibfnamefont {P.}~\bibnamefont {{von Ballmoos}}},\ }\href@noop {}
  {\bibfield  {journal} {\bibinfo  {journal} {Astron. Astrophys.}\ }\textbf
  {\bibinfo {volume} {344}},\ \bibinfo {pages} {68} (\bibinfo {year}
  {1999})}\BibitemShut {NoStop}%
\bibitem [{\citenamefont {{Pl{\"u}schke}}\ \emph {et~al.}(2001)\citenamefont
  {{Pl{\"u}schke}}, \citenamefont {{Diehl}}, \citenamefont {{Sch{\"o}nfelder}},
  \citenamefont {{Bloemen}}, \citenamefont {{Hermsen}}, \citenamefont
  {{Bennett}}, \citenamefont {{Winkler}}, \citenamefont {{McConnell}},
  \citenamefont {{Ryan}}, \citenamefont {{Oberlack}},\ and\ \citenamefont
  {{Kn{\"o}dlseder}}}]{Pluschke2001}%
  \BibitemOpen
  \bibfield  {author} {\bibinfo {author} {\bibfnamefont {S.}~\bibnamefont
  {{Pl{\"u}schke}}}, \bibinfo {author} {\bibfnamefont {R.}~\bibnamefont
  {{Diehl}}}, \bibinfo {author} {\bibfnamefont {V.}~\bibnamefont
  {{Sch{\"o}nfelder}}}, \bibinfo {author} {\bibfnamefont {H.}~\bibnamefont
  {{Bloemen}}}, \bibinfo {author} {\bibfnamefont {W.}~\bibnamefont
  {{Hermsen}}}, \bibinfo {author} {\bibfnamefont {K.}~\bibnamefont
  {{Bennett}}}, \bibinfo {author} {\bibfnamefont {C.}~\bibnamefont
  {{Winkler}}}, \bibinfo {author} {\bibfnamefont {M.}~\bibnamefont
  {{McConnell}}}, \bibinfo {author} {\bibfnamefont {J.}~\bibnamefont {{Ryan}}},
  \bibinfo {author} {\bibfnamefont {U.}~\bibnamefont {{Oberlack}}}, \ and\
  \bibinfo {author} {\bibfnamefont {J.}~\bibnamefont {{Kn{\"o}dlseder}}},\ }in\
  \href@noop {} {\emph {\bibinfo {booktitle} {Exploring the Gamma-Ray
  Universe}}},\ \bibinfo {series} {ESA Special Publication}, Vol.\ \bibinfo
  {volume} {459},\ \bibinfo {editor} {edited by\ \bibinfo {editor}
  {\bibfnamefont {A.}~\bibnamefont {{Gimenez}}}, \bibinfo {editor}
  {\bibfnamefont {V.}~\bibnamefont {{Reglero}}}, \ and\ \bibinfo {editor}
  {\bibfnamefont {C.}~\bibnamefont {{Winkler}}}}\ (\bibinfo {year} {2001})\
  pp.\ \bibinfo {pages} {55--58},\ \Eprint
  {http://arxiv.org/abs/astro-ph/0104047} {astro-ph/0104047} \BibitemShut
  {NoStop}%
\bibitem [{\citenamefont {Smith}(2004)}]{Smith2004}%
  \BibitemOpen
  \bibfield  {author} {\bibinfo {author} {\bibfnamefont {D.}~\bibnamefont
  {Smith}},\ }\href {\doibase https://doi.org/10.1016/j.newar.2003.11.011}
  {\bibfield  {journal} {\bibinfo  {journal} {New Astron. Rev}\ }\textbf
  {\bibinfo {volume} {48}},\ \bibinfo {pages} {87 } (\bibinfo {year}
  {2004})}\BibitemShut {NoStop}%
\bibitem [{\citenamefont {{M. J. Harris}}\ \emph {et~al.}(2005)\citenamefont
  {{M. J. Harris}}, \citenamefont {{J. Kn{\"o}dlseder}}, \citenamefont {{P.
  Jean}}, \citenamefont {{E. Cisana}}, \citenamefont {{R. Diehl}},
  \citenamefont {{G. G. Lichti}}, \citenamefont {{J.-P. Roques}}, \citenamefont
  {{S. Schanne}},\ and\ \citenamefont {{G. Weidenspointner}}}]{Harris2005}%
  \BibitemOpen
  \bibfield  {author} {\bibinfo {author} {\bibnamefont {{M. J. Harris}}},
  \bibinfo {author} {\bibnamefont {{J. Kn{\"o}dlseder}}}, \bibinfo {author}
  {\bibnamefont {{P. Jean}}}, \bibinfo {author} {\bibnamefont {{E. Cisana}}},
  \bibinfo {author} {\bibnamefont {{R. Diehl}}}, \bibinfo {author}
  {\bibnamefont {{G. G. Lichti}}}, \bibinfo {author} {\bibnamefont {{J.-P.
  Roques}}}, \bibinfo {author} {\bibnamefont {{S. Schanne}}}, \ and\ \bibinfo
  {author} {\bibnamefont {{G. Weidenspointner}}},\ }\href {\doibase
  10.1051/0004-6361:200500093} {\bibfield  {journal} {\bibinfo  {journal}
  {{A}stron. {A}strophys.}\ }\textbf {\bibinfo {volume} {433}},\ \bibinfo
  {pages} {L49} (\bibinfo {year} {2005})}\BibitemShut {NoStop}%
\bibitem [{\citenamefont {Rauscher}\ \emph {et~al.}(2002)\citenamefont
  {Rauscher}, \citenamefont {Heger}, \citenamefont {Hoffman},\ and\
  \citenamefont {Woosley}}]{Rauscher2002}%
  \BibitemOpen
  \bibfield  {author} {\bibinfo {author} {\bibfnamefont {T.}~\bibnamefont
  {Rauscher}}, \bibinfo {author} {\bibfnamefont {A.}~\bibnamefont {Heger}},
  \bibinfo {author} {\bibfnamefont {R.~D.}\ \bibnamefont {Hoffman}}, \ and\
  \bibinfo {author} {\bibfnamefont {S.~E.}\ \bibnamefont {Woosley}},\ }\href
  {\doibase 10.1086/341728} {\bibfield  {journal} {\bibinfo  {journal}
  {Astrophys. J.}\ }\textbf {\bibinfo {volume} {576}},\ \bibinfo {pages} {323}
  (\bibinfo {year} {2002})}\BibitemShut {NoStop}%
\bibitem [{\citenamefont {Woosley}\ and\ \citenamefont
  {Heger}(2007)}]{Woosley2007}%
  \BibitemOpen
  \bibfield  {author} {\bibinfo {author} {\bibfnamefont {S.}~\bibnamefont
  {Woosley}}\ and\ \bibinfo {author} {\bibfnamefont {A.}~\bibnamefont
  {Heger}},\ }\href {\doibase https://doi.org/10.1016/j.physrep.2007.02.009}
  {\bibfield  {journal} {\bibinfo  {journal} {Phys. Rep.}\ }\textbf {\bibinfo
  {volume} {442}},\ \bibinfo {pages} {269 } (\bibinfo {year}
  {2007})}\BibitemShut {NoStop}%
\bibitem [{\citenamefont {Parikh}\ \emph {et~al.}(2014)\citenamefont {Parikh},
  \citenamefont {Jos{\'e}},\ and\ \citenamefont {Sala}}]{Parikh2014}%
  \BibitemOpen
  \bibfield  {author} {\bibinfo {author} {\bibfnamefont {A.}~\bibnamefont
  {Parikh}}, \bibinfo {author} {\bibfnamefont {J.}~\bibnamefont {Jos{\'e}}}, \
  and\ \bibinfo {author} {\bibfnamefont {G.}~\bibnamefont {Sala}},\ }\href
  {\doibase 10.1063/1.4863946} {\bibfield  {journal} {\bibinfo  {journal} {AIP
  Advances}\ }\textbf {\bibinfo {volume} {4}},\ \bibinfo {pages} {041002}
  (\bibinfo {year} {2014})}\BibitemShut {NoStop}%
\bibitem [{\citenamefont {Jos{\'{e}}}\ \emph {et~al.}(1997)\citenamefont
  {Jos{\'{e}}}, \citenamefont {Hernanz},\ and\ \citenamefont {Coc}}]{Jose1997}%
  \BibitemOpen
  \bibfield  {author} {\bibinfo {author} {\bibfnamefont {J.}~\bibnamefont
  {Jos{\'{e}}}}, \bibinfo {author} {\bibfnamefont {M.}~\bibnamefont {Hernanz}},
  \ and\ \bibinfo {author} {\bibfnamefont {A.}~\bibnamefont {Coc}},\ }\href
  {\doibase 10.1086/310575} {\bibfield  {journal} {\bibinfo  {journal}
  {Astrophys. J.}\ }\textbf {\bibinfo {volume} {479}},\ \bibinfo {pages} {L55}
  (\bibinfo {year} {1997})}\BibitemShut {NoStop}%
\bibitem [{\citenamefont {Bennett}\ \emph {et~al.}(2013)\citenamefont
  {Bennett}, \citenamefont {Wrede}, \citenamefont {Chipps}, \citenamefont
  {Jos\'e}, \citenamefont {Liddick}, \citenamefont {Santia}, \citenamefont
  {Bowe}, \citenamefont {Chen}, \citenamefont {Cooper}, \citenamefont {Irvine},
  \citenamefont {McNeice}, \citenamefont {Montes}, \citenamefont {Naqvi},
  \citenamefont {Ortez}, \citenamefont {Pain}, \citenamefont {Pereira},
  \citenamefont {Prokop}, \citenamefont {Quaglia}, \citenamefont {Quinn},
  \citenamefont {Schwartz}, \citenamefont {Shanab}, \citenamefont {Simon},
  \citenamefont {Spyrou},\ and\ \citenamefont {Thiagalingam}}]{Bennett2013}%
  \BibitemOpen
  \bibfield  {author} {\bibinfo {author} {\bibfnamefont {M.~B.}\ \bibnamefont
  {Bennett}}, \bibinfo {author} {\bibfnamefont {C.}~\bibnamefont {Wrede}},
  \bibinfo {author} {\bibfnamefont {K.~A.}\ \bibnamefont {Chipps}}, \bibinfo
  {author} {\bibfnamefont {J.}~\bibnamefont {Jos\'e}}, \bibinfo {author}
  {\bibfnamefont {S.~N.}\ \bibnamefont {Liddick}}, \bibinfo {author}
  {\bibfnamefont {M.}~\bibnamefont {Santia}}, \bibinfo {author} {\bibfnamefont
  {A.}~\bibnamefont {Bowe}}, \bibinfo {author} {\bibfnamefont {A.~A.}\
  \bibnamefont {Chen}}, \bibinfo {author} {\bibfnamefont {N.}~\bibnamefont
  {Cooper}}, \bibinfo {author} {\bibfnamefont {D.}~\bibnamefont {Irvine}},
  \bibinfo {author} {\bibfnamefont {E.}~\bibnamefont {McNeice}}, \bibinfo
  {author} {\bibfnamefont {F.}~\bibnamefont {Montes}}, \bibinfo {author}
  {\bibfnamefont {F.}~\bibnamefont {Naqvi}}, \bibinfo {author} {\bibfnamefont
  {R.}~\bibnamefont {Ortez}}, \bibinfo {author} {\bibfnamefont {S.~D.}\
  \bibnamefont {Pain}}, \bibinfo {author} {\bibfnamefont {J.}~\bibnamefont
  {Pereira}}, \bibinfo {author} {\bibfnamefont {C.}~\bibnamefont {Prokop}},
  \bibinfo {author} {\bibfnamefont {J.}~\bibnamefont {Quaglia}}, \bibinfo
  {author} {\bibfnamefont {S.~J.}\ \bibnamefont {Quinn}}, \bibinfo {author}
  {\bibfnamefont {S.~B.}\ \bibnamefont {Schwartz}}, \bibinfo {author}
  {\bibfnamefont {S.}~\bibnamefont {Shanab}}, \bibinfo {author} {\bibfnamefont
  {A.}~\bibnamefont {Simon}}, \bibinfo {author} {\bibfnamefont
  {A.}~\bibnamefont {Spyrou}}, \ and\ \bibinfo {author} {\bibfnamefont
  {E.}~\bibnamefont {Thiagalingam}},\ }\href {\doibase
  10.1103/PhysRevLett.111.232503} {\bibfield  {journal} {\bibinfo  {journal}
  {Phys. Rev. Lett.}\ }\textbf {\bibinfo {volume} {111}},\ \bibinfo {pages}
  {232503} (\bibinfo {year} {2013})}\BibitemShut {NoStop}%
\bibitem [{\citenamefont {Iliadis}\ \emph {et~al.}(1996)\citenamefont
  {Iliadis}, \citenamefont {Buchmann}, \citenamefont {Endt}, \citenamefont
  {Herndl},\ and\ \citenamefont {Wiescher}}]{Iliadis1996}%
  \BibitemOpen
  \bibfield  {author} {\bibinfo {author} {\bibfnamefont {C.}~\bibnamefont
  {Iliadis}}, \bibinfo {author} {\bibfnamefont {L.}~\bibnamefont {Buchmann}},
  \bibinfo {author} {\bibfnamefont {P.~M.}\ \bibnamefont {Endt}}, \bibinfo
  {author} {\bibfnamefont {H.}~\bibnamefont {Herndl}}, \ and\ \bibinfo {author}
  {\bibfnamefont {M.}~\bibnamefont {Wiescher}},\ }\href {\doibase
  10.1103/PhysRevC.53.475} {\bibfield  {journal} {\bibinfo  {journal} {Phys.
  Rev. C}\ }\textbf {\bibinfo {volume} {53}},\ \bibinfo {pages} {475} (\bibinfo
  {year} {1996})}\BibitemShut {NoStop}%
\bibitem [{\citenamefont {Peplowski}\ \emph {et~al.}(2009)\citenamefont
  {Peplowski}, \citenamefont {Baby}, \citenamefont {Wiedenh\"over},
  \citenamefont {Dekat}, \citenamefont {Diffenderfer}, \citenamefont {Gay},
  \citenamefont {Grubor-Urosevic}, \citenamefont {H\"oflich}, \citenamefont
  {Kaye}, \citenamefont {Keeley}, \citenamefont {Rojas},\ and\ \citenamefont
  {Volya}}]{Peplowski2009}%
  \BibitemOpen
  \bibfield  {author} {\bibinfo {author} {\bibfnamefont {P.~N.}\ \bibnamefont
  {Peplowski}}, \bibinfo {author} {\bibfnamefont {L.~T.}\ \bibnamefont {Baby}},
  \bibinfo {author} {\bibfnamefont {I.}~\bibnamefont {Wiedenh\"over}}, \bibinfo
  {author} {\bibfnamefont {S.~E.}\ \bibnamefont {Dekat}}, \bibinfo {author}
  {\bibfnamefont {E.}~\bibnamefont {Diffenderfer}}, \bibinfo {author}
  {\bibfnamefont {D.~L.}\ \bibnamefont {Gay}}, \bibinfo {author} {\bibfnamefont
  {O.}~\bibnamefont {Grubor-Urosevic}}, \bibinfo {author} {\bibfnamefont
  {P.}~\bibnamefont {H\"oflich}}, \bibinfo {author} {\bibfnamefont {R.~A.}\
  \bibnamefont {Kaye}}, \bibinfo {author} {\bibfnamefont {N.}~\bibnamefont
  {Keeley}}, \bibinfo {author} {\bibfnamefont {A.}~\bibnamefont {Rojas}}, \
  and\ \bibinfo {author} {\bibfnamefont {A.}~\bibnamefont {Volya}},\ }\href
  {\doibase 10.1103/PhysRevC.79.032801} {\bibfield  {journal} {\bibinfo
  {journal} {Phys. Rev. C}\ }\textbf {\bibinfo {volume} {79}},\ \bibinfo
  {pages} {032801} (\bibinfo {year} {2009})}\BibitemShut {NoStop}%
\bibitem [{\citenamefont {Basunia}\ and\ \citenamefont
  {Hurst}(2016)}]{datasheet2016}%
  \BibitemOpen
  \bibfield  {author} {\bibinfo {author} {\bibfnamefont {M.}~\bibnamefont
  {Basunia}}\ and\ \bibinfo {author} {\bibfnamefont {A.}~\bibnamefont
  {Hurst}},\ }\href {\doibase https://doi.org/10.1016/j.nds.2016.04.001}
  {\bibfield  {journal} {\bibinfo  {journal} {Nucl. Data Sheets}\ }\textbf
  {\bibinfo {volume} {134}},\ \bibinfo {pages} {1 } (\bibinfo {year}
  {2016})}\BibitemShut {NoStop}%
\bibitem [{\citenamefont {Wang}\ \emph {et~al.}(2017)\citenamefont {Wang},
  \citenamefont {Audi}, \citenamefont {Kondev}, \citenamefont {Huang},
  \citenamefont {Naimi},\ and\ \citenamefont {Xu}}]{AME2016}%
  \BibitemOpen
  \bibfield  {author} {\bibinfo {author} {\bibfnamefont {M.}~\bibnamefont
  {Wang}}, \bibinfo {author} {\bibfnamefont {G.}~\bibnamefont {Audi}}, \bibinfo
  {author} {\bibfnamefont {F.}~\bibnamefont {Kondev}}, \bibinfo {author}
  {\bibfnamefont {W.}~\bibnamefont {Huang}}, \bibinfo {author} {\bibfnamefont
  {S.}~\bibnamefont {Naimi}}, \ and\ \bibinfo {author} {\bibfnamefont
  {X.}~\bibnamefont {Xu}},\ }\href
  {http://stacks.iop.org/1674-1137/41/i=3/a=030003} {\bibfield  {journal}
  {\bibinfo  {journal} {Chin. Phys. C}\ }\textbf {\bibinfo {volume} {41}},\
  \bibinfo {pages} {030003} (\bibinfo {year} {2017})}\BibitemShut {NoStop}%
\bibitem [{\citenamefont {Wrede}(2009)}]{Wrede2009}%
  \BibitemOpen
  \bibfield  {author} {\bibinfo {author} {\bibfnamefont {C.}~\bibnamefont
  {Wrede}},\ }\href {\doibase 10.1103/PhysRevC.79.035803} {\bibfield  {journal}
  {\bibinfo  {journal} {Phys. Rev. C}\ }\textbf {\bibinfo {volume} {79}},\
  \bibinfo {pages} {035803} (\bibinfo {year} {2009})}\BibitemShut {NoStop}%
\bibitem [{\citenamefont {Parpottas}\ \emph {et~al.}(2004)\citenamefont
  {Parpottas}, \citenamefont {Grimes}, \citenamefont {Al-Quraishi},
  \citenamefont {Brune}, \citenamefont {Massey}, \citenamefont {Oldendick},
  \citenamefont {Salas},\ and\ \citenamefont {Wheeler}}]{Parpottas2004}%
  \BibitemOpen
  \bibfield  {author} {\bibinfo {author} {\bibfnamefont {Y.}~\bibnamefont
  {Parpottas}}, \bibinfo {author} {\bibfnamefont {S.~M.}\ \bibnamefont
  {Grimes}}, \bibinfo {author} {\bibfnamefont {S.}~\bibnamefont {Al-Quraishi}},
  \bibinfo {author} {\bibfnamefont {C.~R.}\ \bibnamefont {Brune}}, \bibinfo
  {author} {\bibfnamefont {T.~N.}\ \bibnamefont {Massey}}, \bibinfo {author}
  {\bibfnamefont {J.~E.}\ \bibnamefont {Oldendick}}, \bibinfo {author}
  {\bibfnamefont {A.}~\bibnamefont {Salas}}, \ and\ \bibinfo {author}
  {\bibfnamefont {R.~T.}\ \bibnamefont {Wheeler}},\ }\href {\doibase
  10.1103/PhysRevC.70.065805} {\bibfield  {journal} {\bibinfo  {journal} {Phys.
  Rev. C}\ }\textbf {\bibinfo {volume} {70}},\ \bibinfo {pages} {065805}
  (\bibinfo {year} {2004})}\BibitemShut {NoStop}%
\bibitem [{\citenamefont {Bardayan}\ \emph {et~al.}(2006)\citenamefont
  {Bardayan}, \citenamefont {Howard}, \citenamefont {Blackmon}, \citenamefont
  {Brune}, \citenamefont {Chae}, \citenamefont {Hix}, \citenamefont {Johnson},
  \citenamefont {Jones}, \citenamefont {Kozub}, \citenamefont {Liang},
  \citenamefont {Lingerfelt}, \citenamefont {Livesay}, \citenamefont {Pain},
  \citenamefont {Scott}, \citenamefont {Smith}, \citenamefont {Thomas},\ and\
  \citenamefont {Visser}}]{Bardayan2006}%
  \BibitemOpen
  \bibfield  {author} {\bibinfo {author} {\bibfnamefont {D.~W.}\ \bibnamefont
  {Bardayan}}, \bibinfo {author} {\bibfnamefont {J.~A.}\ \bibnamefont
  {Howard}}, \bibinfo {author} {\bibfnamefont {J.~C.}\ \bibnamefont
  {Blackmon}}, \bibinfo {author} {\bibfnamefont {C.~R.}\ \bibnamefont {Brune}},
  \bibinfo {author} {\bibfnamefont {K.~Y.}\ \bibnamefont {Chae}}, \bibinfo
  {author} {\bibfnamefont {W.~R.}\ \bibnamefont {Hix}}, \bibinfo {author}
  {\bibfnamefont {M.~S.}\ \bibnamefont {Johnson}}, \bibinfo {author}
  {\bibfnamefont {K.~L.}\ \bibnamefont {Jones}}, \bibinfo {author}
  {\bibfnamefont {R.~L.}\ \bibnamefont {Kozub}}, \bibinfo {author}
  {\bibfnamefont {J.~F.}\ \bibnamefont {Liang}}, \bibinfo {author}
  {\bibfnamefont {E.~J.}\ \bibnamefont {Lingerfelt}}, \bibinfo {author}
  {\bibfnamefont {R.~J.}\ \bibnamefont {Livesay}}, \bibinfo {author}
  {\bibfnamefont {S.~D.}\ \bibnamefont {Pain}}, \bibinfo {author}
  {\bibfnamefont {J.~P.}\ \bibnamefont {Scott}}, \bibinfo {author}
  {\bibfnamefont {M.~S.}\ \bibnamefont {Smith}}, \bibinfo {author}
  {\bibfnamefont {J.~S.}\ \bibnamefont {Thomas}}, \ and\ \bibinfo {author}
  {\bibfnamefont {D.~W.}\ \bibnamefont {Visser}},\ }\href {\doibase
  10.1103/PhysRevC.74.045804} {\bibfield  {journal} {\bibinfo  {journal} {Phys.
  Rev. C}\ }\textbf {\bibinfo {volume} {74}},\ \bibinfo {pages} {045804}
  (\bibinfo {year} {2006})}\BibitemShut {NoStop}%
\bibitem [{\citenamefont {Doherty}\ \emph {et~al.}(2015)\citenamefont
  {Doherty}, \citenamefont {Woods}, \citenamefont {Seweryniak}, \citenamefont
  {Albers}, \citenamefont {Ayangeakaa}, \citenamefont {Carpenter},
  \citenamefont {Chiara}, \citenamefont {David}, \citenamefont {Harker},
  \citenamefont {Janssens}, \citenamefont {Kankainen}, \citenamefont
  {Lederer},\ and\ \citenamefont {Zhu}}]{Doherty2015}%
  \BibitemOpen
  \bibfield  {author} {\bibinfo {author} {\bibfnamefont {D.~T.}\ \bibnamefont
  {Doherty}}, \bibinfo {author} {\bibfnamefont {P.~J.}\ \bibnamefont {Woods}},
  \bibinfo {author} {\bibfnamefont {D.}~\bibnamefont {Seweryniak}}, \bibinfo
  {author} {\bibfnamefont {M.}~\bibnamefont {Albers}}, \bibinfo {author}
  {\bibfnamefont {A.~D.}\ \bibnamefont {Ayangeakaa}}, \bibinfo {author}
  {\bibfnamefont {M.~P.}\ \bibnamefont {Carpenter}}, \bibinfo {author}
  {\bibfnamefont {C.~J.}\ \bibnamefont {Chiara}}, \bibinfo {author}
  {\bibfnamefont {H.~M.}\ \bibnamefont {David}}, \bibinfo {author}
  {\bibfnamefont {J.~L.}\ \bibnamefont {Harker}}, \bibinfo {author}
  {\bibfnamefont {R.~V.~F.}\ \bibnamefont {Janssens}}, \bibinfo {author}
  {\bibfnamefont {A.}~\bibnamefont {Kankainen}}, \bibinfo {author}
  {\bibfnamefont {C.}~\bibnamefont {Lederer}}, \ and\ \bibinfo {author}
  {\bibfnamefont {S.}~\bibnamefont {Zhu}},\ }\href {\doibase
  10.1103/PhysRevC.92.035808} {\bibfield  {journal} {\bibinfo  {journal} {Phys.
  Rev. C}\ }\textbf {\bibinfo {volume} {92}},\ \bibinfo {pages} {035808}
  (\bibinfo {year} {2015})}\BibitemShut {NoStop}%
\bibitem [{\citenamefont {Komatsubara}\ \emph {et~al.}(2014)\citenamefont
  {Komatsubara}, \citenamefont {Kubono}, \citenamefont {Hayakawa},
  \citenamefont {Shizuma}, \citenamefont {Ozawa}, \citenamefont {Ito},
  \citenamefont {Ishibashi}, \citenamefont {Moriguchi}, \citenamefont
  {Yamaguchi}, \citenamefont {Kahl}, \citenamefont {Hayakawa}, \citenamefont
  {Nguyen~Binh}, \citenamefont {Chen}, \citenamefont {Chen}, \citenamefont
  {Setoodehnia},\ and\ \citenamefont {Kajino}}]{Komatsubara2014}%
  \BibitemOpen
  \bibfield  {author} {\bibinfo {author} {\bibfnamefont {T.}~\bibnamefont
  {Komatsubara}}, \bibinfo {author} {\bibfnamefont {S.}~\bibnamefont {Kubono}},
  \bibinfo {author} {\bibfnamefont {T.}~\bibnamefont {Hayakawa}}, \bibinfo
  {author} {\bibfnamefont {T.}~\bibnamefont {Shizuma}}, \bibinfo {author}
  {\bibfnamefont {A.}~\bibnamefont {Ozawa}}, \bibinfo {author} {\bibfnamefont
  {Y.}~\bibnamefont {Ito}}, \bibinfo {author} {\bibfnamefont {Y.}~\bibnamefont
  {Ishibashi}}, \bibinfo {author} {\bibfnamefont {T.}~\bibnamefont
  {Moriguchi}}, \bibinfo {author} {\bibfnamefont {H.}~\bibnamefont
  {Yamaguchi}}, \bibinfo {author} {\bibfnamefont {D.}~\bibnamefont {Kahl}},
  \bibinfo {author} {\bibfnamefont {S.}~\bibnamefont {Hayakawa}}, \bibinfo
  {author} {\bibfnamefont {D.}~\bibnamefont {Nguyen~Binh}}, \bibinfo {author}
  {\bibfnamefont {A.~A.}\ \bibnamefont {Chen}}, \bibinfo {author}
  {\bibfnamefont {J.}~\bibnamefont {Chen}}, \bibinfo {author} {\bibfnamefont
  {K.}~\bibnamefont {Setoodehnia}}, \ and\ \bibinfo {author} {\bibfnamefont
  {T.}~\bibnamefont {Kajino}},\ }\href {\doibase 10.1140/epja/i2014-14136-4}
  {\bibfield  {journal} {\bibinfo  {journal} {Eur. Phys. J. A}\ }\textbf
  {\bibinfo {volume} {50}},\ \bibinfo {pages} {136} (\bibinfo {year}
  {2014})}\BibitemShut {NoStop}%
\bibitem [{\citenamefont {Chipps}(2016)}]{Chipps2016}%
  \BibitemOpen
  \bibfield  {author} {\bibinfo {author} {\bibfnamefont {K.~A.}\ \bibnamefont
  {Chipps}},\ }\href {\doibase 10.1103/PhysRevC.93.035801} {\bibfield
  {journal} {\bibinfo  {journal} {Phys. Rev. C}\ }\textbf {\bibinfo {volume}
  {93}},\ \bibinfo {pages} {035801} (\bibinfo {year} {2016})}\BibitemShut
  {NoStop}%
\bibitem [{\citenamefont {P\'erez-Loureiro}\ \emph {et~al.}(2016)\citenamefont
  {P\'erez-Loureiro}, \citenamefont {Wrede}, \citenamefont {Bennett},
  \citenamefont {Liddick}, \citenamefont {Bowe}, \citenamefont {Brown},
  \citenamefont {Chen}, \citenamefont {Chipps}, \citenamefont {Cooper},
  \citenamefont {Irvine}, \citenamefont {McNeice}, \citenamefont {Montes},
  \citenamefont {Naqvi}, \citenamefont {Ortez}, \citenamefont {Pain},
  \citenamefont {Pereira}, \citenamefont {Prokop}, \citenamefont {Quaglia},
  \citenamefont {Quinn}, \citenamefont {Sakstrup}, \citenamefont {Santia},
  \citenamefont {Schwartz}, \citenamefont {Shanab}, \citenamefont {Simon},
  \citenamefont {Spyrou},\ and\ \citenamefont {Thiagalingam}}]{Loureiro2016}%
  \BibitemOpen
  \bibfield  {author} {\bibinfo {author} {\bibfnamefont {D.}~\bibnamefont
  {P\'erez-Loureiro}}, \bibinfo {author} {\bibfnamefont {C.}~\bibnamefont
  {Wrede}}, \bibinfo {author} {\bibfnamefont {M.~B.}\ \bibnamefont {Bennett}},
  \bibinfo {author} {\bibfnamefont {S.~N.}\ \bibnamefont {Liddick}}, \bibinfo
  {author} {\bibfnamefont {A.}~\bibnamefont {Bowe}}, \bibinfo {author}
  {\bibfnamefont {B.~A.}\ \bibnamefont {Brown}}, \bibinfo {author}
  {\bibfnamefont {A.~A.}\ \bibnamefont {Chen}}, \bibinfo {author}
  {\bibfnamefont {K.~A.}\ \bibnamefont {Chipps}}, \bibinfo {author}
  {\bibfnamefont {N.}~\bibnamefont {Cooper}}, \bibinfo {author} {\bibfnamefont
  {D.}~\bibnamefont {Irvine}}, \bibinfo {author} {\bibfnamefont
  {E.}~\bibnamefont {McNeice}}, \bibinfo {author} {\bibfnamefont
  {F.}~\bibnamefont {Montes}}, \bibinfo {author} {\bibfnamefont
  {F.}~\bibnamefont {Naqvi}}, \bibinfo {author} {\bibfnamefont
  {R.}~\bibnamefont {Ortez}}, \bibinfo {author} {\bibfnamefont {S.~D.}\
  \bibnamefont {Pain}}, \bibinfo {author} {\bibfnamefont {J.}~\bibnamefont
  {Pereira}}, \bibinfo {author} {\bibfnamefont {C.~J.}\ \bibnamefont {Prokop}},
  \bibinfo {author} {\bibfnamefont {J.}~\bibnamefont {Quaglia}}, \bibinfo
  {author} {\bibfnamefont {S.~J.}\ \bibnamefont {Quinn}}, \bibinfo {author}
  {\bibfnamefont {J.}~\bibnamefont {Sakstrup}}, \bibinfo {author}
  {\bibfnamefont {M.}~\bibnamefont {Santia}}, \bibinfo {author} {\bibfnamefont
  {S.~B.}\ \bibnamefont {Schwartz}}, \bibinfo {author} {\bibfnamefont
  {S.}~\bibnamefont {Shanab}}, \bibinfo {author} {\bibfnamefont
  {A.}~\bibnamefont {Simon}}, \bibinfo {author} {\bibfnamefont
  {A.}~\bibnamefont {Spyrou}}, \ and\ \bibinfo {author} {\bibfnamefont
  {E.}~\bibnamefont {Thiagalingam}},\ }\href {\doibase
  10.1103/PhysRevC.93.064320} {\bibfield  {journal} {\bibinfo  {journal} {Phys.
  Rev. C}\ }\textbf {\bibinfo {volume} {93}},\ \bibinfo {pages} {064320}
  (\bibinfo {year} {2016})}\BibitemShut {NoStop}%
\bibitem [{\citenamefont {Thomas}\ \emph {et~al.}(2004)\citenamefont {Thomas},
  \citenamefont {Achouri}, \citenamefont {{\"A}yst{\"o}}, \citenamefont
  {B{\'e}raud}, \citenamefont {Blank}, \citenamefont {Canchel}, \citenamefont
  {Czajkowski}, \citenamefont {Dendooven}, \citenamefont {Ensallem},
  \citenamefont {Giovinazzo}, \citenamefont {Guillet}, \citenamefont
  {Honkanen}, \citenamefont {Jokinen}, \citenamefont {Laird}, \citenamefont
  {Lewitowicz}, \citenamefont {Longour}, \citenamefont {de~Oliveira~Santos},
  \citenamefont {Per{\"a}j{\"a}rvi},\ and\ \citenamefont
  {Stanoiu}}]{Thomas2004}%
  \BibitemOpen
  \bibfield  {author} {\bibinfo {author} {\bibfnamefont {J.-C.}\ \bibnamefont
  {Thomas}}, \bibinfo {author} {\bibfnamefont {L.}~\bibnamefont {Achouri}},
  \bibinfo {author} {\bibfnamefont {J.}~\bibnamefont {{\"A}yst{\"o}}}, \bibinfo
  {author} {\bibfnamefont {R.}~\bibnamefont {B{\'e}raud}}, \bibinfo {author}
  {\bibfnamefont {B.}~\bibnamefont {Blank}}, \bibinfo {author} {\bibfnamefont
  {G.}~\bibnamefont {Canchel}}, \bibinfo {author} {\bibfnamefont
  {S.}~\bibnamefont {Czajkowski}}, \bibinfo {author} {\bibfnamefont
  {P.}~\bibnamefont {Dendooven}}, \bibinfo {author} {\bibfnamefont
  {A.}~\bibnamefont {Ensallem}}, \bibinfo {author} {\bibfnamefont
  {J.}~\bibnamefont {Giovinazzo}}, \bibinfo {author} {\bibfnamefont
  {N.}~\bibnamefont {Guillet}}, \bibinfo {author} {\bibfnamefont
  {J.}~\bibnamefont {Honkanen}}, \bibinfo {author} {\bibfnamefont
  {A.}~\bibnamefont {Jokinen}}, \bibinfo {author} {\bibfnamefont
  {A.}~\bibnamefont {Laird}}, \bibinfo {author} {\bibfnamefont
  {M.}~\bibnamefont {Lewitowicz}}, \bibinfo {author} {\bibfnamefont
  {C.}~\bibnamefont {Longour}}, \bibinfo {author} {\bibfnamefont
  {F.}~\bibnamefont {de~Oliveira~Santos}}, \bibinfo {author} {\bibfnamefont
  {K.}~\bibnamefont {Per{\"a}j{\"a}rvi}}, \ and\ \bibinfo {author}
  {\bibfnamefont {M.}~\bibnamefont {Stanoiu}},\ }\href {\doibase
  10.1140/epja/i2003-10218-8} {\bibfield  {journal} {\bibinfo  {journal} {Eur.
  Phys. J. A}\ }\textbf {\bibinfo {volume} {21}},\ \bibinfo {pages} {419}
  (\bibinfo {year} {2004})}\BibitemShut {NoStop}%
\bibitem [{\citenamefont {Janiak}\ \emph {et~al.}(2017)\citenamefont {Janiak},
  \citenamefont {Soko\l{}owska}, \citenamefont {Bezbakh}, \citenamefont
  {Ciemny}, \citenamefont {Czyrkowski}, \citenamefont {D\k{a}browski},
  \citenamefont {Dominik}, \citenamefont {Fomichev}, \citenamefont {Golovkov},
  \citenamefont {Gorshkov}, \citenamefont {Janas}, \citenamefont
  {Kami\ifmmode~\acute{n}\else \'{n}\fi{}ski}, \citenamefont {Knyazev},
  \citenamefont {Krupko}, \citenamefont {Kuich}, \citenamefont {Mazzocchi},
  \citenamefont {Mentel}, \citenamefont {Pf\"utzner}, \citenamefont
  {Pluci\ifmmode~\acute{n}\else \'{n}\fi{}ski}, \citenamefont {Pomorski},
  \citenamefont {Slepniev},\ and\ \citenamefont {Zalewski}}]{Janiak2017}%
  \BibitemOpen
  \bibfield  {author} {\bibinfo {author} {\bibfnamefont {L.}~\bibnamefont
  {Janiak}}, \bibinfo {author} {\bibfnamefont {N.}~\bibnamefont
  {Soko\l{}owska}}, \bibinfo {author} {\bibfnamefont {A.~A.}\ \bibnamefont
  {Bezbakh}}, \bibinfo {author} {\bibfnamefont {A.~A.}\ \bibnamefont {Ciemny}},
  \bibinfo {author} {\bibfnamefont {H.}~\bibnamefont {Czyrkowski}}, \bibinfo
  {author} {\bibfnamefont {R.}~\bibnamefont {D\k{a}browski}}, \bibinfo {author}
  {\bibfnamefont {W.}~\bibnamefont {Dominik}}, \bibinfo {author} {\bibfnamefont
  {A.~S.}\ \bibnamefont {Fomichev}}, \bibinfo {author} {\bibfnamefont {M.~S.}\
  \bibnamefont {Golovkov}}, \bibinfo {author} {\bibfnamefont {A.~V.}\
  \bibnamefont {Gorshkov}}, \bibinfo {author} {\bibfnamefont {Z.}~\bibnamefont
  {Janas}}, \bibinfo {author} {\bibfnamefont {G.}~\bibnamefont
  {Kami\ifmmode~\acute{n}\else \'{n}\fi{}ski}}, \bibinfo {author}
  {\bibfnamefont {A.~G.}\ \bibnamefont {Knyazev}}, \bibinfo {author}
  {\bibfnamefont {S.~A.}\ \bibnamefont {Krupko}}, \bibinfo {author}
  {\bibfnamefont {M.}~\bibnamefont {Kuich}}, \bibinfo {author} {\bibfnamefont
  {C.}~\bibnamefont {Mazzocchi}}, \bibinfo {author} {\bibfnamefont
  {M.}~\bibnamefont {Mentel}}, \bibinfo {author} {\bibfnamefont
  {M.}~\bibnamefont {Pf\"utzner}}, \bibinfo {author} {\bibfnamefont
  {P.}~\bibnamefont {Pluci\ifmmode~\acute{n}\else \'{n}\fi{}ski}}, \bibinfo
  {author} {\bibfnamefont {M.}~\bibnamefont {Pomorski}}, \bibinfo {author}
  {\bibfnamefont {R.~S.}\ \bibnamefont {Slepniev}}, \ and\ \bibinfo {author}
  {\bibfnamefont {B.}~\bibnamefont {Zalewski}},\ }\href {\doibase
  10.1103/PhysRevC.95.034315} {\bibfield  {journal} {\bibinfo  {journal} {Phys.
  Rev. C}\ }\textbf {\bibinfo {volume} {95}},\ \bibinfo {pages} {034315}
  (\bibinfo {year} {2017})}\BibitemShut {NoStop}%
\bibitem [{\citenamefont {Zhan}\ \emph {et~al.}(2008)\citenamefont {Zhan},
  \citenamefont {Xia}, \citenamefont {Zhao}, \citenamefont {Xiao},
  \citenamefont {Yuan}, \citenamefont {Xu}, \citenamefont {Man}, \citenamefont
  {Yuan}, \citenamefont {Gao}, \citenamefont {Yang}, \citenamefont {Song},
  \citenamefont {Cai}, \citenamefont {Yang}, \citenamefont {Sun}, \citenamefont
  {Huang}, \citenamefont {Gan},\ and\ \citenamefont {Wei}}]{Zhan2008533c}%
  \BibitemOpen
  \bibfield  {author} {\bibinfo {author} {\bibfnamefont {W.}~\bibnamefont
  {Zhan}}, \bibinfo {author} {\bibfnamefont {J.}~\bibnamefont {Xia}}, \bibinfo
  {author} {\bibfnamefont {H.}~\bibnamefont {Zhao}}, \bibinfo {author}
  {\bibfnamefont {G.}~\bibnamefont {Xiao}}, \bibinfo {author} {\bibfnamefont
  {Y.}~\bibnamefont {Yuan}}, \bibinfo {author} {\bibfnamefont {H.}~\bibnamefont
  {Xu}}, \bibinfo {author} {\bibfnamefont {K.}~\bibnamefont {Man}}, \bibinfo
  {author} {\bibfnamefont {P.}~\bibnamefont {Yuan}}, \bibinfo {author}
  {\bibfnamefont {D.}~\bibnamefont {Gao}}, \bibinfo {author} {\bibfnamefont
  {X.}~\bibnamefont {Yang}}, \bibinfo {author} {\bibfnamefont {M.}~\bibnamefont
  {Song}}, \bibinfo {author} {\bibfnamefont {X.}~\bibnamefont {Cai}}, \bibinfo
  {author} {\bibfnamefont {X.}~\bibnamefont {Yang}}, \bibinfo {author}
  {\bibfnamefont {Z.}~\bibnamefont {Sun}}, \bibinfo {author} {\bibfnamefont
  {W.}~\bibnamefont {Huang}}, \bibinfo {author} {\bibfnamefont
  {Z.}~\bibnamefont {Gan}}, \ and\ \bibinfo {author} {\bibfnamefont
  {B.}~\bibnamefont {Wei}},\ }\href {\doibase
  https://doi.org/10.1016/j.nuclphysa.2008.02.292} {\bibfield  {journal}
  {\bibinfo  {journal} {Nucl. Phys. A}\ }\textbf {\bibinfo {volume} {805}},\
  \bibinfo {pages} {533c } (\bibinfo {year} {2008})}\BibitemShut {NoStop}%
\bibitem [{\citenamefont {Sun}\ \emph {et~al.}(2003)\citenamefont {Sun},
  \citenamefont {Zhan}, \citenamefont {Guo}, \citenamefont {Xiao},\ and\
  \citenamefont {Li}}]{SUN2003496}%
  \BibitemOpen
  \bibfield  {author} {\bibinfo {author} {\bibfnamefont {Z.}~\bibnamefont
  {Sun}}, \bibinfo {author} {\bibfnamefont {W.-L.}\ \bibnamefont {Zhan}},
  \bibinfo {author} {\bibfnamefont {Z.-Y.}\ \bibnamefont {Guo}}, \bibinfo
  {author} {\bibfnamefont {G.}~\bibnamefont {Xiao}}, \ and\ \bibinfo {author}
  {\bibfnamefont {J.-X.}\ \bibnamefont {Li}},\ }\href {\doibase
  https://doi.org/10.1016/S0168-9002(03)01005-2} {\bibfield  {journal}
  {\bibinfo  {journal} {Nucl. Instrum. Methods Phys. Res., Sect. A}\ }\textbf
  {\bibinfo {volume} {503}},\ \bibinfo {pages} {496 } (\bibinfo {year}
  {2003})}\BibitemShut {NoStop}%
\bibitem [{\citenamefont {Sun}\ \emph {et~al.}(2015)\citenamefont {Sun},
  \citenamefont {Xu}, \citenamefont {Lin}, \citenamefont {Wang}, \citenamefont
  {Fang}, \citenamefont {Li}, \citenamefont {Wang}, \citenamefont {Li},
  \citenamefont {Yang}, \citenamefont {Ma}, \citenamefont {Wang}, \citenamefont
  {Zang}, \citenamefont {Wang}, \citenamefont {Li}, \citenamefont {Shi},
  \citenamefont {Nie}, \citenamefont {Li}, \citenamefont {Li}, \citenamefont
  {Ma}, \citenamefont {Ma}, \citenamefont {Jin}, \citenamefont {Huang},
  \citenamefont {Bai}, \citenamefont {Wang}, \citenamefont {Yang},
  \citenamefont {Jia}, \citenamefont {Zhang}, \citenamefont {Liu},
  \citenamefont {Bao}, \citenamefont {Wang}, \citenamefont {Yang},
  \citenamefont {Zhou}, \citenamefont {Ma},\ and\ \citenamefont
  {Chen}}]{SUN20151}%
  \BibitemOpen
  \bibfield  {author} {\bibinfo {author} {\bibfnamefont {L.}~\bibnamefont
  {Sun}}, \bibinfo {author} {\bibfnamefont {X.}~\bibnamefont {Xu}}, \bibinfo
  {author} {\bibfnamefont {C.}~\bibnamefont {Lin}}, \bibinfo {author}
  {\bibfnamefont {J.}~\bibnamefont {Wang}}, \bibinfo {author} {\bibfnamefont
  {D.}~\bibnamefont {Fang}}, \bibinfo {author} {\bibfnamefont {Z.}~\bibnamefont
  {Li}}, \bibinfo {author} {\bibfnamefont {Y.}~\bibnamefont {Wang}}, \bibinfo
  {author} {\bibfnamefont {J.}~\bibnamefont {Li}}, \bibinfo {author}
  {\bibfnamefont {L.}~\bibnamefont {Yang}}, \bibinfo {author} {\bibfnamefont
  {N.}~\bibnamefont {Ma}}, \bibinfo {author} {\bibfnamefont {K.}~\bibnamefont
  {Wang}}, \bibinfo {author} {\bibfnamefont {H.}~\bibnamefont {Zang}}, \bibinfo
  {author} {\bibfnamefont {H.}~\bibnamefont {Wang}}, \bibinfo {author}
  {\bibfnamefont {C.}~\bibnamefont {Li}}, \bibinfo {author} {\bibfnamefont
  {C.}~\bibnamefont {Shi}}, \bibinfo {author} {\bibfnamefont {M.}~\bibnamefont
  {Nie}}, \bibinfo {author} {\bibfnamefont {X.}~\bibnamefont {Li}}, \bibinfo
  {author} {\bibfnamefont {H.}~\bibnamefont {Li}}, \bibinfo {author}
  {\bibfnamefont {J.}~\bibnamefont {Ma}}, \bibinfo {author} {\bibfnamefont
  {P.}~\bibnamefont {Ma}}, \bibinfo {author} {\bibfnamefont {S.}~\bibnamefont
  {Jin}}, \bibinfo {author} {\bibfnamefont {M.}~\bibnamefont {Huang}}, \bibinfo
  {author} {\bibfnamefont {Z.}~\bibnamefont {Bai}}, \bibinfo {author}
  {\bibfnamefont {J.}~\bibnamefont {Wang}}, \bibinfo {author} {\bibfnamefont
  {F.}~\bibnamefont {Yang}}, \bibinfo {author} {\bibfnamefont {H.}~\bibnamefont
  {Jia}}, \bibinfo {author} {\bibfnamefont {H.}~\bibnamefont {Zhang}}, \bibinfo
  {author} {\bibfnamefont {Z.}~\bibnamefont {Liu}}, \bibinfo {author}
  {\bibfnamefont {P.}~\bibnamefont {Bao}}, \bibinfo {author} {\bibfnamefont
  {D.}~\bibnamefont {Wang}}, \bibinfo {author} {\bibfnamefont {Y.}~\bibnamefont
  {Yang}}, \bibinfo {author} {\bibfnamefont {Y.}~\bibnamefont {Zhou}}, \bibinfo
  {author} {\bibfnamefont {W.}~\bibnamefont {Ma}}, \ and\ \bibinfo {author}
  {\bibfnamefont {J.}~\bibnamefont {Chen}},\ }\href {\doibase
  https://doi.org/10.1016/j.nima.2015.09.039} {\bibfield  {journal} {\bibinfo
  {journal} {Nucl. Instrum. Methods Phys. Res., Sect. A}\ }\textbf {\bibinfo
  {volume} {804}},\ \bibinfo {pages} {1 } (\bibinfo {year} {2015})}\BibitemShut
  {NoStop}%
\bibitem [{\citenamefont {Sun}\ \emph {et~al.}(2019)\citenamefont {Sun},
  \citenamefont {Xu}, \citenamefont {Lin}, \citenamefont {Lee}, \citenamefont
  {Hou}, \citenamefont {Yuan}, \citenamefont {Li}, \citenamefont {Jos\'e},
  \citenamefont {He}, \citenamefont {Wang}, \citenamefont {Wang}, \citenamefont
  {Wu}, \citenamefont {Liang}, \citenamefont {Yang}, \citenamefont {Lam},
  \citenamefont {Ma}, \citenamefont {Duan}, \citenamefont {Gao}, \citenamefont
  {Hu}, \citenamefont {Bai}, \citenamefont {Ma}, \citenamefont {Wang},
  \citenamefont {Zhong}, \citenamefont {Wu}, \citenamefont {Luo}, \citenamefont
  {Jiang}, \citenamefont {Liu}, \citenamefont {Hou}, \citenamefont {Li},
  \citenamefont {Ma}, \citenamefont {Ma}, \citenamefont {Shi}, \citenamefont
  {Yu}, \citenamefont {Patel}, \citenamefont {Jin}, \citenamefont {Wang},
  \citenamefont {Yu}, \citenamefont {Zhou}, \citenamefont {Wang}, \citenamefont
  {Hu}, \citenamefont {Wang}, \citenamefont {Zang}, \citenamefont {Li},
  \citenamefont {Zhao}, \citenamefont {Yang}, \citenamefont {Wen},
  \citenamefont {Yang}, \citenamefont {Jia}, \citenamefont {Zhang},
  \citenamefont {Pan}, \citenamefont {Wang}, \citenamefont {Sun}, \citenamefont
  {Hu}, \citenamefont {Chen}, \citenamefont {Liu}, \citenamefont {Yang},
  \citenamefont {Zhao},\ and\ \citenamefont {Zhang}}]{Sun2019}%
  \BibitemOpen
  \bibfield  {author} {\bibinfo {author} {\bibfnamefont {L.~J.}\ \bibnamefont
  {Sun}}, \bibinfo {author} {\bibfnamefont {X.~X.}\ \bibnamefont {Xu}},
  \bibinfo {author} {\bibfnamefont {C.~J.}\ \bibnamefont {Lin}}, \bibinfo
  {author} {\bibfnamefont {J.}~\bibnamefont {Lee}}, \bibinfo {author}
  {\bibfnamefont {S.~Q.}\ \bibnamefont {Hou}}, \bibinfo {author} {\bibfnamefont
  {C.~X.}\ \bibnamefont {Yuan}}, \bibinfo {author} {\bibfnamefont {Z.~H.}\
  \bibnamefont {Li}}, \bibinfo {author} {\bibfnamefont {J.}~\bibnamefont
  {Jos\'e}}, \bibinfo {author} {\bibfnamefont {J.~J.}\ \bibnamefont {He}},
  \bibinfo {author} {\bibfnamefont {J.~S.}\ \bibnamefont {Wang}}, \bibinfo
  {author} {\bibfnamefont {D.~X.}\ \bibnamefont {Wang}}, \bibinfo {author}
  {\bibfnamefont {H.~Y.}\ \bibnamefont {Wu}}, \bibinfo {author} {\bibfnamefont
  {P.~F.}\ \bibnamefont {Liang}}, \bibinfo {author} {\bibfnamefont {Y.~Y.}\
  \bibnamefont {Yang}}, \bibinfo {author} {\bibfnamefont {Y.~H.}\ \bibnamefont
  {Lam}}, \bibinfo {author} {\bibfnamefont {P.}~\bibnamefont {Ma}}, \bibinfo
  {author} {\bibfnamefont {F.~F.}\ \bibnamefont {Duan}}, \bibinfo {author}
  {\bibfnamefont {Z.~H.}\ \bibnamefont {Gao}}, \bibinfo {author} {\bibfnamefont
  {Q.}~\bibnamefont {Hu}}, \bibinfo {author} {\bibfnamefont {Z.}~\bibnamefont
  {Bai}}, \bibinfo {author} {\bibfnamefont {J.~B.}\ \bibnamefont {Ma}},
  \bibinfo {author} {\bibfnamefont {J.~G.}\ \bibnamefont {Wang}}, \bibinfo
  {author} {\bibfnamefont {F.~P.}\ \bibnamefont {Zhong}}, \bibinfo {author}
  {\bibfnamefont {C.~G.}\ \bibnamefont {Wu}}, \bibinfo {author} {\bibfnamefont
  {D.~W.}\ \bibnamefont {Luo}}, \bibinfo {author} {\bibfnamefont
  {Y.}~\bibnamefont {Jiang}}, \bibinfo {author} {\bibfnamefont
  {Y.}~\bibnamefont {Liu}}, \bibinfo {author} {\bibfnamefont {D.~S.}\
  \bibnamefont {Hou}}, \bibinfo {author} {\bibfnamefont {R.}~\bibnamefont
  {Li}}, \bibinfo {author} {\bibfnamefont {N.~R.}\ \bibnamefont {Ma}}, \bibinfo
  {author} {\bibfnamefont {W.~H.}\ \bibnamefont {Ma}}, \bibinfo {author}
  {\bibfnamefont {G.~Z.}\ \bibnamefont {Shi}}, \bibinfo {author} {\bibfnamefont
  {G.~M.}\ \bibnamefont {Yu}}, \bibinfo {author} {\bibfnamefont
  {D.}~\bibnamefont {Patel}}, \bibinfo {author} {\bibfnamefont {S.~Y.}\
  \bibnamefont {Jin}}, \bibinfo {author} {\bibfnamefont {Y.~F.}\ \bibnamefont
  {Wang}}, \bibinfo {author} {\bibfnamefont {Y.~C.}\ \bibnamefont {Yu}},
  \bibinfo {author} {\bibfnamefont {Q.~W.}\ \bibnamefont {Zhou}}, \bibinfo
  {author} {\bibfnamefont {P.}~\bibnamefont {Wang}}, \bibinfo {author}
  {\bibfnamefont {L.~Y.}\ \bibnamefont {Hu}}, \bibinfo {author} {\bibfnamefont
  {X.}~\bibnamefont {Wang}}, \bibinfo {author} {\bibfnamefont {H.~L.}\
  \bibnamefont {Zang}}, \bibinfo {author} {\bibfnamefont {P.~J.}\ \bibnamefont
  {Li}}, \bibinfo {author} {\bibfnamefont {Q.~Q.}\ \bibnamefont {Zhao}},
  \bibinfo {author} {\bibfnamefont {L.}~\bibnamefont {Yang}}, \bibinfo {author}
  {\bibfnamefont {P.~W.}\ \bibnamefont {Wen}}, \bibinfo {author} {\bibfnamefont
  {F.}~\bibnamefont {Yang}}, \bibinfo {author} {\bibfnamefont {H.~M.}\
  \bibnamefont {Jia}}, \bibinfo {author} {\bibfnamefont {G.~L.}\ \bibnamefont
  {Zhang}}, \bibinfo {author} {\bibfnamefont {M.}~\bibnamefont {Pan}}, \bibinfo
  {author} {\bibfnamefont {X.~Y.}\ \bibnamefont {Wang}}, \bibinfo {author}
  {\bibfnamefont {H.~H.}\ \bibnamefont {Sun}}, \bibinfo {author} {\bibfnamefont
  {Z.~G.}\ \bibnamefont {Hu}}, \bibinfo {author} {\bibfnamefont {R.~F.}\
  \bibnamefont {Chen}}, \bibinfo {author} {\bibfnamefont {M.~L.}\ \bibnamefont
  {Liu}}, \bibinfo {author} {\bibfnamefont {W.~Q.}\ \bibnamefont {Yang}},
  \bibinfo {author} {\bibfnamefont {Y.~M.}\ \bibnamefont {Zhao}}, \ and\
  \bibinfo {author} {\bibfnamefont {H.~Q.}\ \bibnamefont {Zhang}} (\bibinfo
  {collaboration} {RIBLL Collaboration}),\ }\href {\doibase
  10.1103/PhysRevC.99.064312} {\bibfield  {journal} {\bibinfo  {journal} {Phys.
  Rev. C}\ }\textbf {\bibinfo {volume} {99}},\ \bibinfo {pages} {064312}
  (\bibinfo {year} {2019})}\BibitemShut {NoStop}%
\bibitem [{\citenamefont {Li-Jie}\ \emph {et~al.}(2015)\citenamefont {Li-Jie},
  \citenamefont {Cheng-Jian}, \citenamefont {Xin-Xing}, \citenamefont
  {Jian-Song}, \citenamefont {Hui-Ming}, \citenamefont {Feng}, \citenamefont
  {Yan-Yun}, \citenamefont {Lei}, \citenamefont {Peng-Fei}, \citenamefont
  {Huan-Qiao}, \citenamefont {Shi-Lun}, \citenamefont {Zhen-Dong},
  \citenamefont {Ning-Tao}, \citenamefont {Si-Ze}, \citenamefont {Jun-Bing},
  \citenamefont {Peng}, \citenamefont {Nan-Ru},\ and\ \citenamefont
  {Zu-Hua}}]{Sun2015CPL}%
  \BibitemOpen
  \bibfield  {author} {\bibinfo {author} {\bibfnamefont {S.}~\bibnamefont
  {Li-Jie}}, \bibinfo {author} {\bibfnamefont {L.}~\bibnamefont {Cheng-Jian}},
  \bibinfo {author} {\bibfnamefont {X.}~\bibnamefont {Xin-Xing}}, \bibinfo
  {author} {\bibfnamefont {W.}~\bibnamefont {Jian-Song}}, \bibinfo {author}
  {\bibfnamefont {J.}~\bibnamefont {Hui-Ming}}, \bibinfo {author}
  {\bibfnamefont {Y.}~\bibnamefont {Feng}}, \bibinfo {author} {\bibfnamefont
  {Y.}~\bibnamefont {Yan-Yun}}, \bibinfo {author} {\bibfnamefont
  {Y.}~\bibnamefont {Lei}}, \bibinfo {author} {\bibfnamefont {B.}~\bibnamefont
  {Peng-Fei}}, \bibinfo {author} {\bibfnamefont {Z.}~\bibnamefont {Huan-Qiao}},
  \bibinfo {author} {\bibfnamefont {J.}~\bibnamefont {Shi-Lun}}, \bibinfo
  {author} {\bibfnamefont {W.}~\bibnamefont {Zhen-Dong}}, \bibinfo {author}
  {\bibfnamefont {Z.}~\bibnamefont {Ning-Tao}}, \bibinfo {author}
  {\bibfnamefont {C.}~\bibnamefont {Si-Ze}}, \bibinfo {author} {\bibfnamefont
  {M.}~\bibnamefont {Jun-Bing}}, \bibinfo {author} {\bibfnamefont
  {M.}~\bibnamefont {Peng}}, \bibinfo {author} {\bibfnamefont {M.}~\bibnamefont
  {Nan-Ru}}, \ and\ \bibinfo {author} {\bibfnamefont {L.}~\bibnamefont
  {Zu-Hua}},\ }\href {http://stacks.iop.org/0256-307X/32/i=1/a=012301}
  {\bibfield  {journal} {\bibinfo  {journal} {Chin. Phys. Lett.}\ }\textbf
  {\bibinfo {volume} {32}},\ \bibinfo {pages} {012301} (\bibinfo {year}
  {2015})}\BibitemShut {NoStop}%
\bibitem [{\citenamefont {Xu}\ \emph {et~al.}(2017)\citenamefont {Xu},
  \citenamefont {Lin}, \citenamefont {Sun}, \citenamefont {Wang}, \citenamefont
  {Lam}, \citenamefont {Lee}, \citenamefont {Fang}, \citenamefont {Li},
  \citenamefont {Smirnova}, \citenamefont {Yuan}, \citenamefont {Yang},
  \citenamefont {Wang}, \citenamefont {Li}, \citenamefont {Ma}, \citenamefont
  {Wang}, \citenamefont {Zang}, \citenamefont {Wang}, \citenamefont {Li},
  \citenamefont {Liu}, \citenamefont {Wang}, \citenamefont {Shi}, \citenamefont
  {Nie}, \citenamefont {Li}, \citenamefont {Li}, \citenamefont {Ma},
  \citenamefont {Ma}, \citenamefont {Jin}, \citenamefont {Huang}, \citenamefont
  {Bai}, \citenamefont {Yang}, \citenamefont {Jia}, \citenamefont {Liu},
  \citenamefont {Wang}, \citenamefont {Yang}, \citenamefont {Zhou},
  \citenamefont {Ma}, \citenamefont {Chen}, \citenamefont {Hu}, \citenamefont
  {Wang}, \citenamefont {Zhang}, \citenamefont {Ma}, \citenamefont {Zhou},
  \citenamefont {Ma}, \citenamefont {Xu}, \citenamefont {Xiao},\ and\
  \citenamefont {Zhang}}]{Xu2017}%
  \BibitemOpen
  \bibfield  {author} {\bibinfo {author} {\bibfnamefont {X.}~\bibnamefont
  {Xu}}, \bibinfo {author} {\bibfnamefont {C.}~\bibnamefont {Lin}}, \bibinfo
  {author} {\bibfnamefont {L.}~\bibnamefont {Sun}}, \bibinfo {author}
  {\bibfnamefont {J.}~\bibnamefont {Wang}}, \bibinfo {author} {\bibfnamefont
  {Y.}~\bibnamefont {Lam}}, \bibinfo {author} {\bibfnamefont {J.}~\bibnamefont
  {Lee}}, \bibinfo {author} {\bibfnamefont {D.}~\bibnamefont {Fang}}, \bibinfo
  {author} {\bibfnamefont {Z.}~\bibnamefont {Li}}, \bibinfo {author}
  {\bibfnamefont {N.}~\bibnamefont {Smirnova}}, \bibinfo {author}
  {\bibfnamefont {C.}~\bibnamefont {Yuan}}, \bibinfo {author} {\bibfnamefont
  {L.}~\bibnamefont {Yang}}, \bibinfo {author} {\bibfnamefont {Y.}~\bibnamefont
  {Wang}}, \bibinfo {author} {\bibfnamefont {J.}~\bibnamefont {Li}}, \bibinfo
  {author} {\bibfnamefont {N.}~\bibnamefont {Ma}}, \bibinfo {author}
  {\bibfnamefont {K.}~\bibnamefont {Wang}}, \bibinfo {author} {\bibfnamefont
  {H.}~\bibnamefont {Zang}}, \bibinfo {author} {\bibfnamefont {H.}~\bibnamefont
  {Wang}}, \bibinfo {author} {\bibfnamefont {C.}~\bibnamefont {Li}}, \bibinfo
  {author} {\bibfnamefont {M.}~\bibnamefont {Liu}}, \bibinfo {author}
  {\bibfnamefont {J.}~\bibnamefont {Wang}}, \bibinfo {author} {\bibfnamefont
  {C.}~\bibnamefont {Shi}}, \bibinfo {author} {\bibfnamefont {M.}~\bibnamefont
  {Nie}}, \bibinfo {author} {\bibfnamefont {X.}~\bibnamefont {Li}}, \bibinfo
  {author} {\bibfnamefont {H.}~\bibnamefont {Li}}, \bibinfo {author}
  {\bibfnamefont {J.}~\bibnamefont {Ma}}, \bibinfo {author} {\bibfnamefont
  {P.}~\bibnamefont {Ma}}, \bibinfo {author} {\bibfnamefont {S.}~\bibnamefont
  {Jin}}, \bibinfo {author} {\bibfnamefont {M.}~\bibnamefont {Huang}}, \bibinfo
  {author} {\bibfnamefont {Z.}~\bibnamefont {Bai}}, \bibinfo {author}
  {\bibfnamefont {F.}~\bibnamefont {Yang}}, \bibinfo {author} {\bibfnamefont
  {H.}~\bibnamefont {Jia}}, \bibinfo {author} {\bibfnamefont {Z.}~\bibnamefont
  {Liu}}, \bibinfo {author} {\bibfnamefont {D.}~\bibnamefont {Wang}}, \bibinfo
  {author} {\bibfnamefont {Y.}~\bibnamefont {Yang}}, \bibinfo {author}
  {\bibfnamefont {Y.}~\bibnamefont {Zhou}}, \bibinfo {author} {\bibfnamefont
  {W.}~\bibnamefont {Ma}}, \bibinfo {author} {\bibfnamefont {J.}~\bibnamefont
  {Chen}}, \bibinfo {author} {\bibfnamefont {Z.}~\bibnamefont {Hu}}, \bibinfo
  {author} {\bibfnamefont {M.}~\bibnamefont {Wang}}, \bibinfo {author}
  {\bibfnamefont {Y.}~\bibnamefont {Zhang}}, \bibinfo {author} {\bibfnamefont
  {X.}~\bibnamefont {Ma}}, \bibinfo {author} {\bibfnamefont {X.}~\bibnamefont
  {Zhou}}, \bibinfo {author} {\bibfnamefont {Y.}~\bibnamefont {Ma}}, \bibinfo
  {author} {\bibfnamefont {H.}~\bibnamefont {Xu}}, \bibinfo {author}
  {\bibfnamefont {G.}~\bibnamefont {Xiao}}, \ and\ \bibinfo {author}
  {\bibfnamefont {H.}~\bibnamefont {Zhang}},\ }\href {\doibase
  https://doi.org/10.1016/j.physletb.2017.01.028} {\bibfield  {journal}
  {\bibinfo  {journal} {Phys. Lett. B}\ }\textbf {\bibinfo {volume} {766}},\
  \bibinfo {pages} {312 } (\bibinfo {year} {2017})}\BibitemShut {NoStop}%
\bibitem [{\citenamefont {Sun}\ \emph {et~al.}(2017)\citenamefont {Sun},
  \citenamefont {Xu}, \citenamefont {Fang}, \citenamefont {Lin}, \citenamefont
  {Wang}, \citenamefont {Li}, \citenamefont {Wang}, \citenamefont {Li},
  \citenamefont {Yang}, \citenamefont {Ma}, \citenamefont {Wang}, \citenamefont
  {Zang}, \citenamefont {Wang}, \citenamefont {Li}, \citenamefont {Shi},
  \citenamefont {Nie}, \citenamefont {Li}, \citenamefont {Li}, \citenamefont
  {Ma}, \citenamefont {Ma}, \citenamefont {Jin}, \citenamefont {Huang},
  \citenamefont {Bai}, \citenamefont {Wang}, \citenamefont {Yang},
  \citenamefont {Jia}, \citenamefont {Zhang}, \citenamefont {Liu},
  \citenamefont {Bao}, \citenamefont {Wang}, \citenamefont {Yang},
  \citenamefont {Zhou}, \citenamefont {Ma}, \citenamefont {Chen}, \citenamefont
  {Ma}, \citenamefont {Zhang}, \citenamefont {Zhou}, \citenamefont {Xu},
  \citenamefont {Xiao},\ and\ \citenamefont {Zhan}}]{Sun2017}%
  \BibitemOpen
  \bibfield  {author} {\bibinfo {author} {\bibfnamefont {L.~J.}\ \bibnamefont
  {Sun}}, \bibinfo {author} {\bibfnamefont {X.~X.}\ \bibnamefont {Xu}},
  \bibinfo {author} {\bibfnamefont {D.~Q.}\ \bibnamefont {Fang}}, \bibinfo
  {author} {\bibfnamefont {C.~J.}\ \bibnamefont {Lin}}, \bibinfo {author}
  {\bibfnamefont {J.~S.}\ \bibnamefont {Wang}}, \bibinfo {author}
  {\bibfnamefont {Z.~H.}\ \bibnamefont {Li}}, \bibinfo {author} {\bibfnamefont
  {Y.~T.}\ \bibnamefont {Wang}}, \bibinfo {author} {\bibfnamefont
  {J.}~\bibnamefont {Li}}, \bibinfo {author} {\bibfnamefont {L.}~\bibnamefont
  {Yang}}, \bibinfo {author} {\bibfnamefont {N.~R.}\ \bibnamefont {Ma}},
  \bibinfo {author} {\bibfnamefont {K.}~\bibnamefont {Wang}}, \bibinfo {author}
  {\bibfnamefont {H.~L.}\ \bibnamefont {Zang}}, \bibinfo {author}
  {\bibfnamefont {H.~W.}\ \bibnamefont {Wang}}, \bibinfo {author}
  {\bibfnamefont {C.}~\bibnamefont {Li}}, \bibinfo {author} {\bibfnamefont
  {C.~Z.}\ \bibnamefont {Shi}}, \bibinfo {author} {\bibfnamefont {M.~W.}\
  \bibnamefont {Nie}}, \bibinfo {author} {\bibfnamefont {X.~F.}\ \bibnamefont
  {Li}}, \bibinfo {author} {\bibfnamefont {H.}~\bibnamefont {Li}}, \bibinfo
  {author} {\bibfnamefont {J.~B.}\ \bibnamefont {Ma}}, \bibinfo {author}
  {\bibfnamefont {P.}~\bibnamefont {Ma}}, \bibinfo {author} {\bibfnamefont
  {S.~L.}\ \bibnamefont {Jin}}, \bibinfo {author} {\bibfnamefont {M.~R.}\
  \bibnamefont {Huang}}, \bibinfo {author} {\bibfnamefont {Z.}~\bibnamefont
  {Bai}}, \bibinfo {author} {\bibfnamefont {J.~G.}\ \bibnamefont {Wang}},
  \bibinfo {author} {\bibfnamefont {F.}~\bibnamefont {Yang}}, \bibinfo {author}
  {\bibfnamefont {H.~M.}\ \bibnamefont {Jia}}, \bibinfo {author} {\bibfnamefont
  {H.~Q.}\ \bibnamefont {Zhang}}, \bibinfo {author} {\bibfnamefont {Z.~H.}\
  \bibnamefont {Liu}}, \bibinfo {author} {\bibfnamefont {P.~F.}\ \bibnamefont
  {Bao}}, \bibinfo {author} {\bibfnamefont {D.~X.}\ \bibnamefont {Wang}},
  \bibinfo {author} {\bibfnamefont {Y.~Y.}\ \bibnamefont {Yang}}, \bibinfo
  {author} {\bibfnamefont {Y.~J.}\ \bibnamefont {Zhou}}, \bibinfo {author}
  {\bibfnamefont {W.~H.}\ \bibnamefont {Ma}}, \bibinfo {author} {\bibfnamefont
  {J.}~\bibnamefont {Chen}}, \bibinfo {author} {\bibfnamefont {Y.~G.}\
  \bibnamefont {Ma}}, \bibinfo {author} {\bibfnamefont {Y.~H.}\ \bibnamefont
  {Zhang}}, \bibinfo {author} {\bibfnamefont {X.~H.}\ \bibnamefont {Zhou}},
  \bibinfo {author} {\bibfnamefont {H.~S.}\ \bibnamefont {Xu}}, \bibinfo
  {author} {\bibfnamefont {G.~Q.}\ \bibnamefont {Xiao}}, \ and\ \bibinfo
  {author} {\bibfnamefont {W.~L.}\ \bibnamefont {Zhan}},\ }\href {\doibase
  10.1103/PhysRevC.95.014314} {\bibfield  {journal} {\bibinfo  {journal} {Phys.
  Rev. C}\ }\textbf {\bibinfo {volume} {95}},\ \bibinfo {pages} {014314}
  (\bibinfo {year} {2017})}\BibitemShut {NoStop}%
\bibitem [{\citenamefont {Wang}\ \emph
  {et~al.}(2018{\natexlab{a}})\citenamefont {Wang}, \citenamefont {Fang},
  \citenamefont {Wang}, \citenamefont {Xu}, \citenamefont {Sun}, \citenamefont
  {Bai}, \citenamefont {Huang}, \citenamefont {Jin}, \citenamefont {Li},
  \citenamefont {Li}, \citenamefont {Li}, \citenamefont {Li}, \citenamefont
  {Lin}, \citenamefont {Ma}, \citenamefont {Ma}, \citenamefont {Ma},
  \citenamefont {Nie}, \citenamefont {Shi}, \citenamefont {Wang}, \citenamefont
  {Wang}, \citenamefont {Wang}, \citenamefont {Yang}, \citenamefont {Yang},
  \citenamefont {Zhang}, \citenamefont {Zhou}, \citenamefont {Ma},\ and\
  \citenamefont {Shen}}]{WangK2018}%
  \BibitemOpen
  \bibfield  {author} {\bibinfo {author} {\bibfnamefont {K.}~\bibnamefont
  {Wang}}, \bibinfo {author} {\bibfnamefont {D.~Q.}\ \bibnamefont {Fang}},
  \bibinfo {author} {\bibfnamefont {Y.~T.}\ \bibnamefont {Wang}}, \bibinfo
  {author} {\bibfnamefont {X.~X.}\ \bibnamefont {Xu}}, \bibinfo {author}
  {\bibfnamefont {L.~J.}\ \bibnamefont {Sun}}, \bibinfo {author} {\bibfnamefont
  {Z.}~\bibnamefont {Bai}}, \bibinfo {author} {\bibfnamefont {M.~R.}\
  \bibnamefont {Huang}}, \bibinfo {author} {\bibfnamefont {S.~L.}\ \bibnamefont
  {Jin}}, \bibinfo {author} {\bibfnamefont {C.}~\bibnamefont {Li}}, \bibinfo
  {author} {\bibfnamefont {H.}~\bibnamefont {Li}}, \bibinfo {author}
  {\bibfnamefont {J.}~\bibnamefont {Li}}, \bibinfo {author} {\bibfnamefont
  {X.~F.}\ \bibnamefont {Li}}, \bibinfo {author} {\bibfnamefont {C.~J.}\
  \bibnamefont {Lin}}, \bibinfo {author} {\bibfnamefont {J.~B.}\ \bibnamefont
  {Ma}}, \bibinfo {author} {\bibfnamefont {P.}~\bibnamefont {Ma}}, \bibinfo
  {author} {\bibfnamefont {W.~H.}\ \bibnamefont {Ma}}, \bibinfo {author}
  {\bibfnamefont {M.~W.}\ \bibnamefont {Nie}}, \bibinfo {author} {\bibfnamefont
  {C.~Z.}\ \bibnamefont {Shi}}, \bibinfo {author} {\bibfnamefont {H.~W.}\
  \bibnamefont {Wang}}, \bibinfo {author} {\bibfnamefont {J.~G.}\ \bibnamefont
  {Wang}}, \bibinfo {author} {\bibfnamefont {J.~S.}\ \bibnamefont {Wang}},
  \bibinfo {author} {\bibfnamefont {L.}~\bibnamefont {Yang}}, \bibinfo {author}
  {\bibfnamefont {Y.~Y.}\ \bibnamefont {Yang}}, \bibinfo {author}
  {\bibfnamefont {H.~Q.}\ \bibnamefont {Zhang}}, \bibinfo {author}
  {\bibfnamefont {Y.~J.}\ \bibnamefont {Zhou}}, \bibinfo {author}
  {\bibfnamefont {Y.~G.}\ \bibnamefont {Ma}}, \ and\ \bibinfo {author}
  {\bibfnamefont {W.~Q.}\ \bibnamefont {Shen}},\ }\href {\doibase
  10.1142/S0218301318500143} {\bibfield  {journal} {\bibinfo  {journal} {Int.
  J. Mod. Phys. E}\ }\textbf {\bibinfo {volume} {27}},\ \bibinfo {pages}
  {1850014} (\bibinfo {year} {2018}{\natexlab{a}})}\BibitemShut {NoStop}%
\bibitem [{\citenamefont {Wang}\ \emph
  {et~al.}(2018{\natexlab{b}})\citenamefont {Wang}, \citenamefont {Fang},
  \citenamefont {Wang}, \citenamefont {Xu}, \citenamefont {Sun}, \citenamefont
  {Bai}, \citenamefont {Bao}, \citenamefont {Cao}, \citenamefont {Dai},
  \citenamefont {Ding}, \citenamefont {He}, \citenamefont {Huang},
  \citenamefont {Jin}, \citenamefont {Li}, \citenamefont {Lin}, \citenamefont
  {Liu}, \citenamefont {Lv}, \citenamefont {Ma}, \citenamefont {Ma},
  \citenamefont {Wang}, \citenamefont {Wang}, \citenamefont {Wang},
  \citenamefont {Wang}, \citenamefont {Yang}, \citenamefont {Ye}, \citenamefont
  {Zhang}, \citenamefont {Zhao}, \citenamefont {Zhou}, \citenamefont {Ma},\
  and\ \citenamefont {Shen}}]{WangYT20181}%
  \BibitemOpen
  \bibfield  {author} {\bibinfo {author} {\bibfnamefont {Y.-T.}\ \bibnamefont
  {Wang}}, \bibinfo {author} {\bibfnamefont {D.-Q.}\ \bibnamefont {Fang}},
  \bibinfo {author} {\bibfnamefont {K.}~\bibnamefont {Wang}}, \bibinfo {author}
  {\bibfnamefont {X.-X.}\ \bibnamefont {Xu}}, \bibinfo {author} {\bibfnamefont
  {L.-J.}\ \bibnamefont {Sun}}, \bibinfo {author} {\bibfnamefont
  {Z.}~\bibnamefont {Bai}}, \bibinfo {author} {\bibfnamefont {P.-F.}\
  \bibnamefont {Bao}}, \bibinfo {author} {\bibfnamefont {X.-G.}\ \bibnamefont
  {Cao}}, \bibinfo {author} {\bibfnamefont {Z.-T.}\ \bibnamefont {Dai}},
  \bibinfo {author} {\bibfnamefont {B.}~\bibnamefont {Ding}}, \bibinfo {author}
  {\bibfnamefont {W.-B.}\ \bibnamefont {He}}, \bibinfo {author} {\bibfnamefont
  {M.-R.}\ \bibnamefont {Huang}}, \bibinfo {author} {\bibfnamefont {S.-L.}\
  \bibnamefont {Jin}}, \bibinfo {author} {\bibfnamefont {Y.}~\bibnamefont
  {Li}}, \bibinfo {author} {\bibfnamefont {C.-J.}\ \bibnamefont {Lin}},
  \bibinfo {author} {\bibfnamefont {L.-X.}\ \bibnamefont {Liu}}, \bibinfo
  {author} {\bibfnamefont {M.}~\bibnamefont {Lv}}, \bibinfo {author}
  {\bibfnamefont {J.-B.}\ \bibnamefont {Ma}}, \bibinfo {author} {\bibfnamefont
  {P.}~\bibnamefont {Ma}}, \bibinfo {author} {\bibfnamefont {H.-W.}\
  \bibnamefont {Wang}}, \bibinfo {author} {\bibfnamefont {J.-G.}\ \bibnamefont
  {Wang}}, \bibinfo {author} {\bibfnamefont {J.-S.}\ \bibnamefont {Wang}},
  \bibinfo {author} {\bibfnamefont {S.-T.}\ \bibnamefont {Wang}}, \bibinfo
  {author} {\bibfnamefont {Y.-Y.}\ \bibnamefont {Yang}}, \bibinfo {author}
  {\bibfnamefont {S.-Q.}\ \bibnamefont {Ye}}, \bibinfo {author} {\bibfnamefont
  {H.-Q.}\ \bibnamefont {Zhang}}, \bibinfo {author} {\bibfnamefont {M.-H.}\
  \bibnamefont {Zhao}}, \bibinfo {author} {\bibfnamefont {C.-L.}\ \bibnamefont
  {Zhou}}, \bibinfo {author} {\bibfnamefont {Y.-G.}\ \bibnamefont {Ma}}, \ and\
  \bibinfo {author} {\bibfnamefont {W.-Q.}\ \bibnamefont {Shen}},\ }\href
  {\doibase 10.1140/epja/i2018-12543-1} {\bibfield  {journal} {\bibinfo
  {journal} {Eur. Phys. J. A}\ }\textbf {\bibinfo {volume} {54}},\ \bibinfo
  {pages} {107} (\bibinfo {year} {2018}{\natexlab{b}})}\BibitemShut {NoStop}%
\bibitem [{\citenamefont {Xu}\ \emph {et~al.}(2018)\citenamefont {Xu},
  \citenamefont {Teh}, \citenamefont {Lin}, \citenamefont {Lee}, \citenamefont
  {Yang}, \citenamefont {Guo}, \citenamefont {Guo}, \citenamefont {Sun},
  \citenamefont {Teng}, \citenamefont {Liu}, \citenamefont {Li}, \citenamefont
  {Liang}, \citenamefont {Yang}, \citenamefont {Ma}, \citenamefont {Jia},
  \citenamefont {Wang}, \citenamefont {Leblond}, \citenamefont {Lokotko},
  \citenamefont {Zhao},\ and\ \citenamefont {Zhang}}]{Xu2018}%
  \BibitemOpen
  \bibfield  {author} {\bibinfo {author} {\bibfnamefont {X.-X.}\ \bibnamefont
  {Xu}}, \bibinfo {author} {\bibfnamefont {F.~C.~E.}\ \bibnamefont {Teh}},
  \bibinfo {author} {\bibfnamefont {C.-J.}\ \bibnamefont {Lin}}, \bibinfo
  {author} {\bibfnamefont {J.}~\bibnamefont {Lee}}, \bibinfo {author}
  {\bibfnamefont {F.}~\bibnamefont {Yang}}, \bibinfo {author} {\bibfnamefont
  {Z.-Q.}\ \bibnamefont {Guo}}, \bibinfo {author} {\bibfnamefont {T.-S.}\
  \bibnamefont {Guo}}, \bibinfo {author} {\bibfnamefont {L.-J.}\ \bibnamefont
  {Sun}}, \bibinfo {author} {\bibfnamefont {X.-Z.}\ \bibnamefont {Teng}},
  \bibinfo {author} {\bibfnamefont {J.-J.}\ \bibnamefont {Liu}}, \bibinfo
  {author} {\bibfnamefont {P.-J.}\ \bibnamefont {Li}}, \bibinfo {author}
  {\bibfnamefont {P.-F.}\ \bibnamefont {Liang}}, \bibinfo {author}
  {\bibfnamefont {L.}~\bibnamefont {Yang}}, \bibinfo {author} {\bibfnamefont
  {N.-R.}\ \bibnamefont {Ma}}, \bibinfo {author} {\bibfnamefont {H.-M.}\
  \bibnamefont {Jia}}, \bibinfo {author} {\bibfnamefont {D.-X.}\ \bibnamefont
  {Wang}}, \bibinfo {author} {\bibfnamefont {S.}~\bibnamefont {Leblond}},
  \bibinfo {author} {\bibfnamefont {T.}~\bibnamefont {Lokotko}}, \bibinfo
  {author} {\bibfnamefont {Q.-Q.}\ \bibnamefont {Zhao}}, \ and\ \bibinfo
  {author} {\bibfnamefont {H.-Q.}\ \bibnamefont {Zhang}},\ }\href {\doibase
  10.1007/s41365-018-0406-0} {\bibfield  {journal} {\bibinfo  {journal} {Nucl.
  Sci. Technol.}\ }\textbf {\bibinfo {volume} {29}},\ \bibinfo {pages} {73}
  (\bibinfo {year} {2018})}\BibitemShut {NoStop}%
\bibitem [{\citenamefont {G{\"o}rres}\ \emph {et~al.}(1992)\citenamefont
  {G{\"o}rres}, \citenamefont {Wiescher}, \citenamefont {Scheller},
  \citenamefont {Morrissey}, \citenamefont {Sherrill}, \citenamefont {Bazin},\
  and\ \citenamefont {Winger}}]{Gorres1992}%
  \BibitemOpen
  \bibfield  {author} {\bibinfo {author} {\bibfnamefont {J.}~\bibnamefont
  {G{\"o}rres}}, \bibinfo {author} {\bibfnamefont {M.}~\bibnamefont
  {Wiescher}}, \bibinfo {author} {\bibfnamefont {K.}~\bibnamefont {Scheller}},
  \bibinfo {author} {\bibfnamefont {D.~J.}\ \bibnamefont {Morrissey}}, \bibinfo
  {author} {\bibfnamefont {B.~M.}\ \bibnamefont {Sherrill}}, \bibinfo {author}
  {\bibfnamefont {D.}~\bibnamefont {Bazin}}, \ and\ \bibinfo {author}
  {\bibfnamefont {J.~A.}\ \bibnamefont {Winger}},\ }\href {\doibase
  10.1103/PhysRevC.46.R833} {\bibfield  {journal} {\bibinfo  {journal} {Phys.
  Rev. C}\ }\textbf {\bibinfo {volume} {46}},\ \bibinfo {pages} {R833}
  (\bibinfo {year} {1992})}\BibitemShut {NoStop}%
\bibitem [{\citenamefont {Trinder}\ \emph {et~al.}(1997)\citenamefont
  {Trinder}, \citenamefont {Adelberger}, \citenamefont {Brown}, \citenamefont
  {Janas}, \citenamefont {Keller}, \citenamefont {Krumbholz}, \citenamefont
  {Kunze}, \citenamefont {Magnus}, \citenamefont {Meissner}, \citenamefont
  {Piechaczek}, \citenamefont {Pf{\"u}tzner}, \citenamefont {Roeckl},
  \citenamefont {Rykaczewski}, \citenamefont {Schmidt-Ott},\ and\ \citenamefont
  {Weber}}]{TRINDER1997}%
  \BibitemOpen
  \bibfield  {author} {\bibinfo {author} {\bibfnamefont {W.}~\bibnamefont
  {Trinder}}, \bibinfo {author} {\bibfnamefont {E.}~\bibnamefont {Adelberger}},
  \bibinfo {author} {\bibfnamefont {B.}~\bibnamefont {Brown}}, \bibinfo
  {author} {\bibfnamefont {Z.}~\bibnamefont {Janas}}, \bibinfo {author}
  {\bibfnamefont {H.}~\bibnamefont {Keller}}, \bibinfo {author} {\bibfnamefont
  {K.}~\bibnamefont {Krumbholz}}, \bibinfo {author} {\bibfnamefont
  {V.}~\bibnamefont {Kunze}}, \bibinfo {author} {\bibfnamefont
  {P.}~\bibnamefont {Magnus}}, \bibinfo {author} {\bibfnamefont
  {F.}~\bibnamefont {Meissner}}, \bibinfo {author} {\bibfnamefont
  {A.}~\bibnamefont {Piechaczek}}, \bibinfo {author} {\bibfnamefont
  {M.}~\bibnamefont {Pf{\"u}tzner}}, \bibinfo {author} {\bibfnamefont
  {E.}~\bibnamefont {Roeckl}}, \bibinfo {author} {\bibfnamefont
  {K.}~\bibnamefont {Rykaczewski}}, \bibinfo {author} {\bibfnamefont {W.-D.}\
  \bibnamefont {Schmidt-Ott}}, \ and\ \bibinfo {author} {\bibfnamefont
  {M.}~\bibnamefont {Weber}},\ }\href {\doibase
  https://doi.org/10.1016/S0375-9474(97)00163-2} {\bibfield  {journal}
  {\bibinfo  {journal} {Nucl. Phys. A}\ }\textbf {\bibinfo {volume} {620}},\
  \bibinfo {pages} {191 } (\bibinfo {year} {1997})}\BibitemShut {NoStop}%
\bibitem [{\citenamefont {Achouri}\ \emph {et~al.}(2006)\citenamefont
  {Achouri}, \citenamefont {de~Oliveira~Santos}, \citenamefont {Lewitowicz},
  \citenamefont {Blank}, \citenamefont {{\"A}yst{\"o}}, \citenamefont
  {Canchel}, \citenamefont {Czajkowski}, \citenamefont {Dendooven},
  \citenamefont {Emsallem}, \citenamefont {Giovinazzo}, \citenamefont
  {Guillet}, \citenamefont {Jokinen}, \citenamefont {Laird}, \citenamefont
  {Longour}, \citenamefont {Per{\"a}j{\"a}rvi}, \citenamefont {Smirnova},
  \citenamefont {Stanoiu},\ and\ \citenamefont {Thomas}}]{Achouri2006}%
  \BibitemOpen
  \bibfield  {author} {\bibinfo {author} {\bibfnamefont {N.~L.}\ \bibnamefont
  {Achouri}}, \bibinfo {author} {\bibfnamefont {F.}~\bibnamefont
  {de~Oliveira~Santos}}, \bibinfo {author} {\bibfnamefont {M.}~\bibnamefont
  {Lewitowicz}}, \bibinfo {author} {\bibfnamefont {B.}~\bibnamefont {Blank}},
  \bibinfo {author} {\bibfnamefont {J.}~\bibnamefont {{\"A}yst{\"o}}}, \bibinfo
  {author} {\bibfnamefont {G.}~\bibnamefont {Canchel}}, \bibinfo {author}
  {\bibfnamefont {S.}~\bibnamefont {Czajkowski}}, \bibinfo {author}
  {\bibfnamefont {P.}~\bibnamefont {Dendooven}}, \bibinfo {author}
  {\bibfnamefont {A.}~\bibnamefont {Emsallem}}, \bibinfo {author}
  {\bibfnamefont {J.}~\bibnamefont {Giovinazzo}}, \bibinfo {author}
  {\bibfnamefont {N.}~\bibnamefont {Guillet}}, \bibinfo {author} {\bibfnamefont
  {A.}~\bibnamefont {Jokinen}}, \bibinfo {author} {\bibfnamefont {A.~M.}\
  \bibnamefont {Laird}}, \bibinfo {author} {\bibfnamefont {C.}~\bibnamefont
  {Longour}}, \bibinfo {author} {\bibfnamefont {K.}~\bibnamefont
  {Per{\"a}j{\"a}rvi}}, \bibinfo {author} {\bibfnamefont {N.}~\bibnamefont
  {Smirnova}}, \bibinfo {author} {\bibfnamefont {M.}~\bibnamefont {Stanoiu}}, \
  and\ \bibinfo {author} {\bibfnamefont {J.~C.}\ \bibnamefont {Thomas}},\
  }\href {\doibase 10.1140/epja/i2005-10274-0} {\bibfield  {journal} {\bibinfo
  {journal} {Eur. Phys. J. A}\ }\textbf {\bibinfo {volume} {27}},\ \bibinfo
  {pages} {287} (\bibinfo {year} {2006})}\BibitemShut {NoStop}%
\bibitem [{\citenamefont {Wildenthal}(1984)}]{Wildenthal1984}%
  \BibitemOpen
  \bibfield  {author} {\bibinfo {author} {\bibfnamefont {B.}~\bibnamefont
  {Wildenthal}},\ }\href {\doibase
  https://doi.org/10.1016/0146-6410(84)90011-5} {\bibfield  {journal} {\bibinfo
   {journal} {Prog. Part. Nucl. Phys.}\ }\textbf {\bibinfo {volume} {11}},\
  \bibinfo {pages} {5 } (\bibinfo {year} {1984})}\BibitemShut {NoStop}%
\bibitem [{\citenamefont {Brown}\ and\ \citenamefont
  {Richter}(2006)}]{Brown2006}%
  \BibitemOpen
  \bibfield  {author} {\bibinfo {author} {\bibfnamefont {B.~A.}\ \bibnamefont
  {Brown}}\ and\ \bibinfo {author} {\bibfnamefont {W.~A.}\ \bibnamefont
  {Richter}},\ }\href {\doibase 10.1103/PhysRevC.74.034315} {\bibfield
  {journal} {\bibinfo  {journal} {Phys. Rev. C}\ }\textbf {\bibinfo {volume}
  {74}},\ \bibinfo {pages} {034315} (\bibinfo {year} {2006})}\BibitemShut
  {NoStop}%
\bibitem [{\citenamefont {Brown}\ and\ \citenamefont
  {Wildenthal}(1988)}]{Brown1988}%
  \BibitemOpen
  \bibfield  {author} {\bibinfo {author} {\bibfnamefont {B.~A.}\ \bibnamefont
  {Brown}}\ and\ \bibinfo {author} {\bibfnamefont {B.~H.}\ \bibnamefont
  {Wildenthal}},\ }\href {\doibase 10.1146/annurev.ns.38.120188.000333}
  {\bibfield  {journal} {\bibinfo  {journal} {Annu. Rev. Nucl. Part. Sci.}\
  }\textbf {\bibinfo {volume} {38}},\ \bibinfo {pages} {29} (\bibinfo {year}
  {1988})}\BibitemShut {NoStop}%
\bibitem [{\citenamefont {Yuan}\ \emph {et~al.}(2014)\citenamefont {Yuan},
  \citenamefont {Qi}, \citenamefont {Xu}, \citenamefont {Suzuki},\ and\
  \citenamefont {Otsuka}}]{Yuan2014}%
  \BibitemOpen
  \bibfield  {author} {\bibinfo {author} {\bibfnamefont {C.}~\bibnamefont
  {Yuan}}, \bibinfo {author} {\bibfnamefont {C.}~\bibnamefont {Qi}}, \bibinfo
  {author} {\bibfnamefont {F.}~\bibnamefont {Xu}}, \bibinfo {author}
  {\bibfnamefont {T.}~\bibnamefont {Suzuki}}, \ and\ \bibinfo {author}
  {\bibfnamefont {T.}~\bibnamefont {Otsuka}},\ }\href {\doibase
  10.1103/PhysRevC.89.044327} {\bibfield  {journal} {\bibinfo  {journal} {Phys.
  Rev. C}\ }\textbf {\bibinfo {volume} {89}},\ \bibinfo {pages} {044327}
  (\bibinfo {year} {2014})}\BibitemShut {NoStop}%
\bibitem [{\citenamefont {Yuan}\ \emph {et~al.}(2012)\citenamefont {Yuan},
  \citenamefont {Suzuki}, \citenamefont {Otsuka}, \citenamefont {Xu},\ and\
  \citenamefont {Tsunoda}}]{Yuan2012}%
  \BibitemOpen
  \bibfield  {author} {\bibinfo {author} {\bibfnamefont {C.}~\bibnamefont
  {Yuan}}, \bibinfo {author} {\bibfnamefont {T.}~\bibnamefont {Suzuki}},
  \bibinfo {author} {\bibfnamefont {T.}~\bibnamefont {Otsuka}}, \bibinfo
  {author} {\bibfnamefont {F.}~\bibnamefont {Xu}}, \ and\ \bibinfo {author}
  {\bibfnamefont {N.}~\bibnamefont {Tsunoda}},\ }\href {\doibase
  10.1103/PhysRevC.85.064324} {\bibfield  {journal} {\bibinfo  {journal} {Phys.
  Rev. C}\ }\textbf {\bibinfo {volume} {85}},\ \bibinfo {pages} {064324}
  (\bibinfo {year} {2012})}\BibitemShut {NoStop}%
\bibitem [{\citenamefont {Utsuno}\ \emph {et~al.}(1999)\citenamefont {Utsuno},
  \citenamefont {Otsuka}, \citenamefont {Mizusaki},\ and\ \citenamefont
  {Honma}}]{Utsuno1999}%
  \BibitemOpen
  \bibfield  {author} {\bibinfo {author} {\bibfnamefont {Y.}~\bibnamefont
  {Utsuno}}, \bibinfo {author} {\bibfnamefont {T.}~\bibnamefont {Otsuka}},
  \bibinfo {author} {\bibfnamefont {T.}~\bibnamefont {Mizusaki}}, \ and\
  \bibinfo {author} {\bibfnamefont {M.}~\bibnamefont {Honma}},\ }\href
  {\doibase 10.1103/PhysRevC.60.054315} {\bibfield  {journal} {\bibinfo
  {journal} {Phys. Rev. C}\ }\textbf {\bibinfo {volume} {60}},\ \bibinfo
  {pages} {054315} (\bibinfo {year} {1999})}\BibitemShut {NoStop}%
\bibitem [{\citenamefont {Richter}\ \emph {et~al.}(2011)\citenamefont
  {Richter}, \citenamefont {Brown}, \citenamefont {Signoracci},\ and\
  \citenamefont {Wiescher}}]{Richter2011}%
  \BibitemOpen
  \bibfield  {author} {\bibinfo {author} {\bibfnamefont {W.~A.}\ \bibnamefont
  {Richter}}, \bibinfo {author} {\bibfnamefont {B.~A.}\ \bibnamefont {Brown}},
  \bibinfo {author} {\bibfnamefont {A.}~\bibnamefont {Signoracci}}, \ and\
  \bibinfo {author} {\bibfnamefont {M.}~\bibnamefont {Wiescher}},\ }\href
  {\doibase 10.1103/PhysRevC.83.065803} {\bibfield  {journal} {\bibinfo
  {journal} {Phys. Rev. C}\ }\textbf {\bibinfo {volume} {83}},\ \bibinfo
  {pages} {065803} (\bibinfo {year} {2011})}\BibitemShut {NoStop}%
\bibitem [{\citenamefont {Iliadis}(2008)}]{Iliadis2008}%
  \BibitemOpen
  \bibfield  {author} {\bibinfo {author} {\bibfnamefont {C.}~\bibnamefont
  {Iliadis}},\ }\href@noop {} {\emph {\bibinfo {title} {Nuclear Physics of
  Stars}}}\ (\bibinfo  {publisher} {Wiley},\ \bibinfo {year}
  {2008})\BibitemShut {NoStop}%
\bibitem [{\citenamefont {Rolfs}\ and\ \citenamefont
  {Rodney}(1988)}]{Rolfs1988}%
  \BibitemOpen
  \bibfield  {author} {\bibinfo {author} {\bibfnamefont {C.~E.}\ \bibnamefont
  {Rolfs}}\ and\ \bibinfo {author} {\bibfnamefont {W.~S.}\ \bibnamefont
  {Rodney}},\ }\href@noop {} {\emph {\bibinfo {title} {Cauldrons in the
  Cosmos}}}\ (\bibinfo  {publisher} {University of Chicago},\ \bibinfo {year}
  {1988})\BibitemShut {NoStop}%
\bibitem [{\citenamefont {He}\ \emph {et~al.}(2017)\citenamefont {He},
  \citenamefont {Parikh}, \citenamefont {Xu}, \citenamefont {Zhang},
  \citenamefont {Zhou},\ and\ \citenamefont {Xu}}]{He2017}%
  \BibitemOpen
  \bibfield  {author} {\bibinfo {author} {\bibfnamefont {J.~J.}\ \bibnamefont
  {He}}, \bibinfo {author} {\bibfnamefont {A.}~\bibnamefont {Parikh}}, \bibinfo
  {author} {\bibfnamefont {Y.}~\bibnamefont {Xu}}, \bibinfo {author}
  {\bibfnamefont {Y.~H.}\ \bibnamefont {Zhang}}, \bibinfo {author}
  {\bibfnamefont {X.~H.}\ \bibnamefont {Zhou}}, \ and\ \bibinfo {author}
  {\bibfnamefont {H.~S.}\ \bibnamefont {Xu}},\ }\href {\doibase
  10.1103/PhysRevC.96.045801} {\bibfield  {journal} {\bibinfo  {journal} {Phys.
  Rev. C}\ }\textbf {\bibinfo {volume} {96}},\ \bibinfo {pages} {045801}
  (\bibinfo {year} {2017})}\BibitemShut {NoStop}%
\bibitem [{\citenamefont {Herndl}\ \emph {et~al.}(1995)\citenamefont {Herndl},
  \citenamefont {G{\"o}rres}, \citenamefont {Wiescher}, \citenamefont {Brown},\
  and\ \citenamefont {Van~Wormer}}]{Herndl1995}%
  \BibitemOpen
  \bibfield  {author} {\bibinfo {author} {\bibfnamefont {H.}~\bibnamefont
  {Herndl}}, \bibinfo {author} {\bibfnamefont {J.}~\bibnamefont {G{\"o}rres}},
  \bibinfo {author} {\bibfnamefont {M.}~\bibnamefont {Wiescher}}, \bibinfo
  {author} {\bibfnamefont {B.~A.}\ \bibnamefont {Brown}}, \ and\ \bibinfo
  {author} {\bibfnamefont {L.}~\bibnamefont {Van~Wormer}},\ }\href {\doibase
  10.1103/PhysRevC.52.1078} {\bibfield  {journal} {\bibinfo  {journal} {Phys.
  Rev. C}\ }\textbf {\bibinfo {volume} {52}},\ \bibinfo {pages} {1078}
  (\bibinfo {year} {1995})}\BibitemShut {NoStop}%
\bibitem [{\citenamefont {Fowler}\ \emph {et~al.}(1967)\citenamefont {Fowler},
  \citenamefont {Caughlan},\ and\ \citenamefont {Zimmerman}}]{Fowler1967}%
  \BibitemOpen
  \bibfield  {author} {\bibinfo {author} {\bibfnamefont {W.~A.}\ \bibnamefont
  {Fowler}}, \bibinfo {author} {\bibfnamefont {G.~R.}\ \bibnamefont
  {Caughlan}}, \ and\ \bibinfo {author} {\bibfnamefont {B.~A.}\ \bibnamefont
  {Zimmerman}},\ }\href {\doibase 10.1146/annurev.aa.05.090167.002521}
  {\bibfield  {journal} {\bibinfo  {journal} {Annu. Rev. Astron. Astrophys.}\
  }\textbf {\bibinfo {volume} {5}},\ \bibinfo {pages} {525} (\bibinfo {year}
  {1967})}\BibitemShut {NoStop}%
\bibitem [{\citenamefont {Matic}\ \emph {et~al.}(2010)\citenamefont {Matic},
  \citenamefont {van~den Berg}, \citenamefont {Harakeh}, \citenamefont
  {W{\"o}rtche}, \citenamefont {Berg}, \citenamefont {Couder}, \citenamefont
  {G\"orres}, \citenamefont {LeBlanc}, \citenamefont {O'Brien}, \citenamefont
  {Wiescher}, \citenamefont {Fujita}, \citenamefont {Hatanaka}, \citenamefont
  {Sakemi}, \citenamefont {Shimizu}, \citenamefont {Tameshige}, \citenamefont
  {Tamii}, \citenamefont {Yosoi}, \citenamefont {Adachi}, \citenamefont
  {Fujita}, \citenamefont {Shimbara}, \citenamefont {Fujita}, \citenamefont
  {Wakasa}, \citenamefont {Brown},\ and\ \citenamefont {Schatz}}]{Matic2010}%
  \BibitemOpen
  \bibfield  {author} {\bibinfo {author} {\bibfnamefont {A.}~\bibnamefont
  {Matic}}, \bibinfo {author} {\bibfnamefont {A.~M.}\ \bibnamefont {van~den
  Berg}}, \bibinfo {author} {\bibfnamefont {M.~N.}\ \bibnamefont {Harakeh}},
  \bibinfo {author} {\bibfnamefont {H.~J.}\ \bibnamefont {W{\"o}rtche}},
  \bibinfo {author} {\bibfnamefont {G.~P.~A.}\ \bibnamefont {Berg}}, \bibinfo
  {author} {\bibfnamefont {M.}~\bibnamefont {Couder}}, \bibinfo {author}
  {\bibfnamefont {J.}~\bibnamefont {G\"orres}}, \bibinfo {author}
  {\bibfnamefont {P.}~\bibnamefont {LeBlanc}}, \bibinfo {author} {\bibfnamefont
  {S.}~\bibnamefont {O'Brien}}, \bibinfo {author} {\bibfnamefont
  {M.}~\bibnamefont {Wiescher}}, \bibinfo {author} {\bibfnamefont
  {K.}~\bibnamefont {Fujita}}, \bibinfo {author} {\bibfnamefont
  {K.}~\bibnamefont {Hatanaka}}, \bibinfo {author} {\bibfnamefont
  {Y.}~\bibnamefont {Sakemi}}, \bibinfo {author} {\bibfnamefont
  {Y.}~\bibnamefont {Shimizu}}, \bibinfo {author} {\bibfnamefont
  {Y.}~\bibnamefont {Tameshige}}, \bibinfo {author} {\bibfnamefont
  {A.}~\bibnamefont {Tamii}}, \bibinfo {author} {\bibfnamefont
  {M.}~\bibnamefont {Yosoi}}, \bibinfo {author} {\bibfnamefont
  {T.}~\bibnamefont {Adachi}}, \bibinfo {author} {\bibfnamefont
  {Y.}~\bibnamefont {Fujita}}, \bibinfo {author} {\bibfnamefont
  {Y.}~\bibnamefont {Shimbara}}, \bibinfo {author} {\bibfnamefont
  {H.}~\bibnamefont {Fujita}}, \bibinfo {author} {\bibfnamefont
  {T.}~\bibnamefont {Wakasa}}, \bibinfo {author} {\bibfnamefont {B.~A.}\
  \bibnamefont {Brown}}, \ and\ \bibinfo {author} {\bibfnamefont
  {H.}~\bibnamefont {Schatz}},\ }\href {\doibase 10.1103/PhysRevC.82.025807}
  {\bibfield  {journal} {\bibinfo  {journal} {Phys. Rev. C}\ }\textbf {\bibinfo
  {volume} {82}},\ \bibinfo {pages} {025807} (\bibinfo {year}
  {2010})}\BibitemShut {NoStop}%
\bibitem [{\citenamefont {Iliadis}\ \emph {et~al.}(2001)\citenamefont
  {Iliadis}, \citenamefont {D'Auria}, \citenamefont {Starrfield}, \citenamefont
  {Thompson},\ and\ \citenamefont {Wiescher}}]{Iliadis2001}%
  \BibitemOpen
  \bibfield  {author} {\bibinfo {author} {\bibfnamefont {C.}~\bibnamefont
  {Iliadis}}, \bibinfo {author} {\bibfnamefont {J.~M.}\ \bibnamefont
  {D'Auria}}, \bibinfo {author} {\bibfnamefont {S.}~\bibnamefont {Starrfield}},
  \bibinfo {author} {\bibfnamefont {W.~J.}\ \bibnamefont {Thompson}}, \ and\
  \bibinfo {author} {\bibfnamefont {M.}~\bibnamefont {Wiescher}},\ }\href
  {http://stacks.iop.org/0067-0049/134/i=1/a=151} {\bibfield  {journal}
  {\bibinfo  {journal} {Astrophys. J. Suppl. Ser.}\ }\textbf {\bibinfo {volume}
  {134}},\ \bibinfo {pages} {151} (\bibinfo {year} {2001})}\BibitemShut
  {NoStop}%
\bibitem [{\citenamefont {Cyburt}\ \emph {et~al.}(2010)\citenamefont {Cyburt},
  \citenamefont {Amthor}, \citenamefont {Ferguson}, \citenamefont {Meisel},
  \citenamefont {Smith}, \citenamefont {Warren}, \citenamefont {Heger},
  \citenamefont {Hoffman}, \citenamefont {Rauscher}, \citenamefont {Sakharuk},
  \citenamefont {Schatz}, \citenamefont {Thielemann},\ and\ \citenamefont
  {Wiescher}}]{Cyburt2010}%
  \BibitemOpen
  \bibfield  {author} {\bibinfo {author} {\bibfnamefont {R.~H.}\ \bibnamefont
  {Cyburt}}, \bibinfo {author} {\bibfnamefont {A.~M.}\ \bibnamefont {Amthor}},
  \bibinfo {author} {\bibfnamefont {R.}~\bibnamefont {Ferguson}}, \bibinfo
  {author} {\bibfnamefont {Z.}~\bibnamefont {Meisel}}, \bibinfo {author}
  {\bibfnamefont {K.}~\bibnamefont {Smith}}, \bibinfo {author} {\bibfnamefont
  {S.}~\bibnamefont {Warren}}, \bibinfo {author} {\bibfnamefont
  {A.}~\bibnamefont {Heger}}, \bibinfo {author} {\bibfnamefont {R.~D.}\
  \bibnamefont {Hoffman}}, \bibinfo {author} {\bibfnamefont {T.}~\bibnamefont
  {Rauscher}}, \bibinfo {author} {\bibfnamefont {A.}~\bibnamefont {Sakharuk}},
  \bibinfo {author} {\bibfnamefont {H.}~\bibnamefont {Schatz}}, \bibinfo
  {author} {\bibfnamefont {F.~K.}\ \bibnamefont {Thielemann}}, \ and\ \bibinfo
  {author} {\bibfnamefont {M.}~\bibnamefont {Wiescher}},\ }\href
  {http://stacks.iop.org/0067-0049/189/i=1/a=240} {\bibfield  {journal}
  {\bibinfo  {journal} {Astrophys. J. Suppl. Ser.}\ }\textbf {\bibinfo {volume}
  {189}},\ \bibinfo {pages} {240} (\bibinfo {year} {2010})}\BibitemShut
  {NoStop}%
\bibitem [{\citenamefont {{Prantzos}}\ and\ \citenamefont
  {{Cass{\'e}}}(1986)}]{Prantzos1986}%
  \BibitemOpen
  \bibfield  {author} {\bibinfo {author} {\bibfnamefont {N.}~\bibnamefont
  {{Prantzos}}}\ and\ \bibinfo {author} {\bibfnamefont {M.}~\bibnamefont
  {{Cass{\'e}}}},\ }\href {\doibase 10.1086/164419} {\bibfield  {journal}
  {\bibinfo  {journal} {\apj}\ }\textbf {\bibinfo {volume} {307}},\ \bibinfo
  {pages} {324} (\bibinfo {year} {1986})}\BibitemShut {NoStop}%
\bibitem [{\citenamefont {{Arnett}}\ and\ \citenamefont
  {{Wefel}}(1978)}]{Arnett1978}%
  \BibitemOpen
  \bibfield  {author} {\bibinfo {author} {\bibfnamefont {W.~D.}\ \bibnamefont
  {{Arnett}}}\ and\ \bibinfo {author} {\bibfnamefont {J.~P.}\ \bibnamefont
  {{Wefel}}},\ }\href {\doibase 10.1086/182778} {\bibfield  {journal} {\bibinfo
   {journal} {\apj}\ }\textbf {\bibinfo {volume} {224}},\ \bibinfo {pages}
  {L139} (\bibinfo {year} {1978})}\BibitemShut {NoStop}%
\bibitem [{\citenamefont {{Woosley}}\ and\ \citenamefont
  {{Weaver}}(1980)}]{Woosley1980}%
  \BibitemOpen
  \bibfield  {author} {\bibinfo {author} {\bibfnamefont {S.~E.}\ \bibnamefont
  {{Woosley}}}\ and\ \bibinfo {author} {\bibfnamefont {T.~A.}\ \bibnamefont
  {{Weaver}}},\ }\href {\doibase 10.1086/158067} {\bibfield  {journal}
  {\bibinfo  {journal} {\apj}\ }\textbf {\bibinfo {volume} {238}},\ \bibinfo
  {pages} {1017} (\bibinfo {year} {1980})}\BibitemShut {NoStop}%
\bibitem [{\citenamefont {{Norgaard}}(1980)}]{Norgaard1980}%
  \BibitemOpen
  \bibfield  {author} {\bibinfo {author} {\bibfnamefont {H.}~\bibnamefont
  {{Norgaard}}},\ }\href {\doibase 10.1086/157815} {\bibfield  {journal}
  {\bibinfo  {journal} {\apj}\ }\textbf {\bibinfo {volume} {236}},\ \bibinfo
  {pages} {895} (\bibinfo {year} {1980})}\BibitemShut {NoStop}%
\bibitem [{\citenamefont {{Forestini}}\ \emph {et~al.}(1991)\citenamefont
  {{Forestini}}, \citenamefont {{Arnould}},\ and\ \citenamefont
  {{Paulus}}}]{Forestini1991}%
  \BibitemOpen
  \bibfield  {author} {\bibinfo {author} {\bibfnamefont {M.}~\bibnamefont
  {{Forestini}}}, \bibinfo {author} {\bibfnamefont {M.}~\bibnamefont
  {{Arnould}}}, \ and\ \bibinfo {author} {\bibfnamefont {G.}~\bibnamefont
  {{Paulus}}},\ }\href@noop {} {\bibfield  {journal} {\bibinfo  {journal}
  {Astron. Astrophys.}\ }\textbf {\bibinfo {volume} {252}},\ \bibinfo {pages}
  {597} (\bibinfo {year} {1991})}\BibitemShut {NoStop}%
\bibitem [{\citenamefont {Starrfield}\ \emph {et~al.}(1993)\citenamefont
  {Starrfield}, \citenamefont {Truran}, \citenamefont {Politano}, \citenamefont
  {Sparks}, \citenamefont {Nofar},\ and\ \citenamefont
  {Shaviv}}]{Starrfield1993}%
  \BibitemOpen
  \bibfield  {author} {\bibinfo {author} {\bibfnamefont {S.}~\bibnamefont
  {Starrfield}}, \bibinfo {author} {\bibfnamefont {J.}~\bibnamefont {Truran}},
  \bibinfo {author} {\bibfnamefont {M.}~\bibnamefont {Politano}}, \bibinfo
  {author} {\bibfnamefont {W.}~\bibnamefont {Sparks}}, \bibinfo {author}
  {\bibfnamefont {I.}~\bibnamefont {Nofar}}, \ and\ \bibinfo {author}
  {\bibfnamefont {G.}~\bibnamefont {Shaviv}},\ }\href {\doibase
  https://doi.org/10.1016/0370-1573(93)90067-N} {\bibfield  {journal} {\bibinfo
   {journal} {Phys. Rep.}\ }\textbf {\bibinfo {volume} {227}},\ \bibinfo
  {pages} {223 } (\bibinfo {year} {1993})}\BibitemShut {NoStop}%
\bibitem [{\citenamefont {{Jos{\'e}}}\ and\ \citenamefont
  {{Hernanz}}(1998)}]{Jose1998}%
  \BibitemOpen
  \bibfield  {author} {\bibinfo {author} {\bibfnamefont {J.}~\bibnamefont
  {{Jos{\'e}}}}\ and\ \bibinfo {author} {\bibfnamefont {M.}~\bibnamefont
  {{Hernanz}}},\ }\href {\doibase 10.1086/305244} {\bibfield  {journal}
  {\bibinfo  {journal} {\apj}\ }\textbf {\bibinfo {volume} {494}},\ \bibinfo
  {pages} {680} (\bibinfo {year} {1998})}\BibitemShut {NoStop}%
\bibitem [{\citenamefont {Shimizu}\ \emph {et~al.}(2019)\citenamefont
  {Shimizu}, \citenamefont {Mizusaki}, \citenamefont {Utsuno},\ and\
  \citenamefont {Tsunoda}}]{Shimizu2019}%
  \BibitemOpen
  \bibfield  {author} {\bibinfo {author} {\bibfnamefont {N.}~\bibnamefont
  {Shimizu}}, \bibinfo {author} {\bibfnamefont {T.}~\bibnamefont {Mizusaki}},
  \bibinfo {author} {\bibfnamefont {Y.}~\bibnamefont {Utsuno}}, \ and\ \bibinfo
  {author} {\bibfnamefont {Y.}~\bibnamefont {Tsunoda}},\ }\href {\doibase
  https://doi.org/10.1016/j.cpc.2019.06.011} {\bibfield  {journal} {\bibinfo
  {journal} {Comput. Phys. Commun.}\ }\textbf {\bibinfo {volume} {244}},\
  \bibinfo {pages} {372 } (\bibinfo {year} {2019})}\BibitemShut {NoStop}%
\end{thebibliography}%

\end{document}